\documentclass[acmsmall,table,xcdraw,dvipsnames]{acmart}

\usepackage{algorithmic}
\usepackage{algorithm}
\usepackage{textcomp}
\usepackage{stfloats}
\usepackage{url}
\usepackage{verbatim}
\usepackage{graphicx}

\usepackage{subcaption}

\usepackage{amsmath}
\usepackage{multirow}
\usepackage{multicol}
\usepackage[normalem]{ulem}
\useunder{\uline}{\ul}{}
\usepackage{balance}
\usepackage{makecell}
\usepackage[marginal]{footmisc}

\usepackage{varwidth}
\usepackage{caption}
\usepackage{graphicx}
\usepackage{float} 
\usepackage{marvosym}
\usepackage{bm}
\usepackage{enumitem}
\usepackage{colortbl} 
\usepackage{subcaption}
\usepackage[most]{tcolorbox}

\AtBeginDocument{%
  }

\newtcolorbox{summarybox}{
    colback=gray!10!white,  
    colframe=gray!60!black, 
    boxrule=0.8pt,          
    arc=2mm,                
    left=1mm, right=1mm, top=1mm, bottom=1mm,
    fonttitle=\bfseries,
    before skip=8pt, after skip=8pt
}
\newcommand{\stitle}[1]{\vspace{1ex} \noindent{\bf #1}}

\setcopyright{acmlicensed}
\copyrightyear{2018}
\acmYear{2018}
\acmDOI{XXXXXXX.XXXXXXX}
\acmConference[Conference acronym 'XX]{Make sure to enter the correct
  conference title from your rights confirmation email}{June 03--05,
  2018}{Woodstock, NY}
\acmISBN{978-1-4503-XXXX-X/2018/06}

\begin{document}

\title{When Graph Contrastive Learning Backfires: Spectral Vulnerability and Defense in Recommendation}

\author{Zongwei Wang}
\email{zongwei@cqu.edu.cn}
\affiliation{%
  \institution{Chongqing University}
  \city{Chongqing}
  \country{China}
}

\author{Min Gao}
\email{gaomin@cqu.edu.cn}
\authornote{Corresponding author}
\affiliation{%
  \institution{Key Laboratory of Dependable Service Computing in Cyber Physical Society (Chongqing University), Ministry of Education}
  \city{Chongqing}
  \country{China}}

\author{Junliang Yu}
\affiliation{%
  \institution{The University of Queensland}
  \city{Brisbane}
  \country{Australia}}
\email{jl.yu@uq.edu.au}


\author{Shazia Sadiq}
\email{shazia@itee.uq.edu.au}
\affiliation{%
  \institution{The University of Queensland}
    \city{Brisbane}
  \country{Australia}}
  
\author{Hongzhi Yin}
\email{h.yin1@uq.edu.au}
\affiliation{%
  \institution{The University of Queensland}
    \city{Brisbane}
  \country{Australia}}

\author{Ling Liu}
\email{lingliu@cc.gatech.edu}
\affiliation{%
  \institution{Georgia Institute of Technology}
  \city{Atlanta}
  \country{USA}}

\renewcommand{\shortauthors}{Wang, Gao, et al.}

\begin{abstract}
Graph Contrastive Learning (GCL) has demonstrated substantial promise in enhancing the robustness and generalization of recommender systems, particularly by enabling models to leverage large-scale unlabeled data for improved representation learning. However, in this paper, we reveal an unexpected vulnerability: the integration of GCL inadvertently increases the susceptibility of a recommender to targeted promotion attacks. Through both theoretical investigation and empirical validation, we identify the root cause as the spectral smoothing effect induced by contrastive optimization, which disperses item embeddings across the representation space and unintentionally enhances the exposure of target items. Building on this insight, we introduce CLeaR, a bi-level optimization attack method that deliberately amplifies spectral smoothness, enabling a systematic investigation of the susceptibility of GCL-based recommendation models to targeted promotion attacks. Our findings highlight the urgent need for robust countermeasures; in response, we further propose SIM, a spectral irregularity mitigation framework designed to accurately detect and suppress targeted items without compromising model performance. Extensive experiments on multiple benchmark datasets demonstrate that, compared to existing targeted promotion attacks, GCL-based recommendation models exhibit greater susceptibility when evaluated with CLeaR, while SIM effectively mitigates these vulnerabilities.
\end{abstract}




\maketitle

\section{Introduction}
Graph Contrastive Learning (GCL) \cite{27jaiswal2020survey, 28khosla2020supervised} facilitates a self-supervised learning paradigm to extract robust representations from unlabeled large-scale datasets, thereby enhancing the generalization and accuracy of the model in various computational tasks\cite{35you2020graph, 36shuai2022review}. In the context of recommender systems, extensive empirical studies have shown that incorporating GCL not only enhances the robustness of the recommendation but also sustains the performance of the system under noisy or adversarial conditions \cite{06yu2023self,25yu2022xsimgcl}. 

However, current robustness evaluations of GCL-based recommendation frameworks primarily focus on system-wide performance metrics, often overlooking vulnerabilities at the level of individual items. For example, on e-commerce platforms, adversaries can exploit recommendation pipelines to artificially increase the visibility of low-quality products for financial gain \cite{04wang2022gray}, while in news recommendation ecosystems, targeted misinformation campaigns can manipulate exposure rates for specific user cohorts \cite{30yi2022ua}. These vulnerabilities stem from \textbf{targeted promotion attacks} \cite{13zhang2022pipattack, 69yuan2023manipulating}, wherein attackers inject crafted malicious interactions to disrupt recommendation outputs. Given the widespread occurrence and potential impact of targeted promotion attacks in real-world scenarios, we raise a critical question: \textit{Can GCL-based recommendation frameworks effectively withstand the risks posed by targeted promotion attacks?}

\begin{figure}[t]
    \centering
    \begin{subfigure}[b]{0.98\textwidth}
        \centering
        \includegraphics[width=\textwidth]{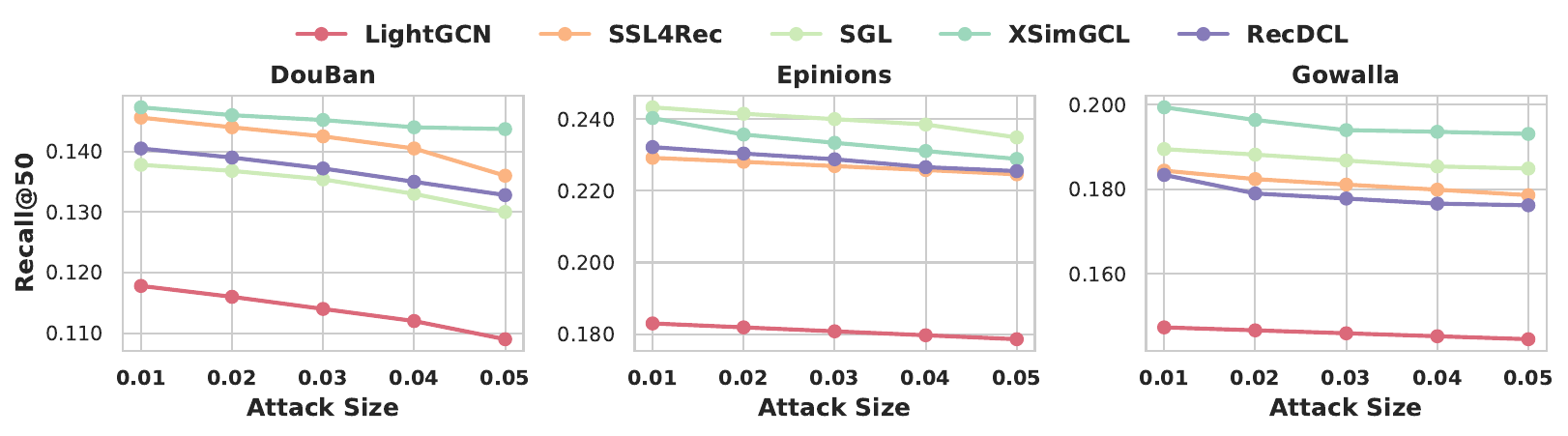}
        \caption{Recommendation performance comparison (Recall@50).}
        \label{fig:image1}
    \end{subfigure}    
    \hfill
    \begin{subfigure}[b]{0.98\textwidth}
        \centering
        \includegraphics[width=\textwidth]{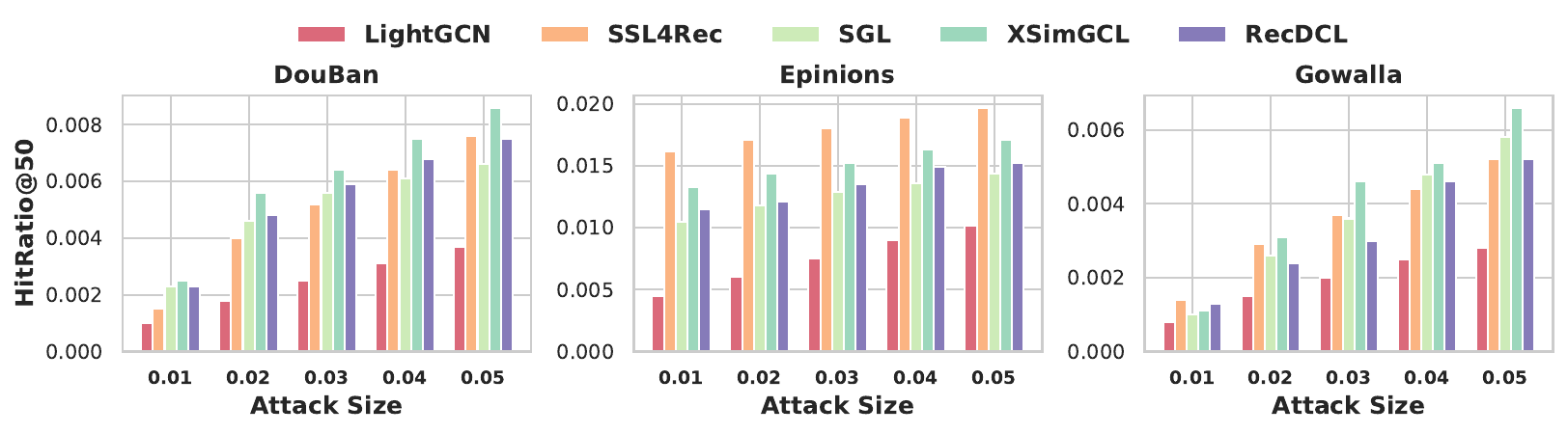}
        \caption{Attack performance comparison (Hit Ratio@50).}
        \label{fig:image2}
    \end{subfigure}

    \caption{The comparison of LightGCN, SSL4Rec, XSimGCL, and RecDCL on DouBan, Epinions, and Gowalla datasets under RandomAttack. Attack size represents the ratio of the number of malicious users to the total number of users.}
    \label{introduction}
    \vspace{-10pt}
\end{figure}

To address this concern, we began by evaluating how GCL techniques influence individual items in recommender systems. Specifically, we benchmarked LightGCN \cite{07he2020lightgcn} against four representative GCL-enhanced models, including SSL4Rec \cite{55yao2021self}, SGL \cite{08wu2021self}, XSimGCL \cite{25yu2022xsimgcl}, and RecDCL \cite{72zhang2024recdcl}. Using the standard RandomAttack strategy \cite{10lam2004shilling}, we injected synthetic attacker profiles into public datasets and measured two metrics: overall recommendation quality via Recall@50, and the attackers’ success at promoting specific items via Hit Ratio@50. As shown in Fig.~\ref{introduction}, although the GCL-based methods maintained their global recommendation accuracy, they consistently exhibited greater gains in the promotion of poisoned target items compared to LightGCN alone. These surprising findings reveal that, while GCL models are designed for robust representation learning, they may inadvertently increase the exposure of specific target items, raising serious concerns about the potential for malicious exploitation in GCL-based recommender systems.

To further investigate the root cause of GCL’s unexpected susceptibility to targeted promotion attacks, we conduct both theoretical analyses and empirical evaluations, demonstrating that the InfoNCE objective function \cite{37oord2018representation} plays a critical role in increasing the exposure of target items. Specifically, the optimization process in GCL encourages a smoother distribution of spectral values in the learned representations, resulting in a more dispersed embedding space compared to non-GCL methods. This phenomenon makes GCL a double-edged sword: while it mitigates popularity bias by suppressing overly popular items \cite{26abdollahpouri2019popularity}, it also facilitates the inclusion of less popular items in recommendation lists. Since targeted promotion attacks often focus on such unpopular items, GCL inadvertently introduces a vulnerability by naturally increasing the exposure of these targets.

Motivated by these insights, our prior work \cite{80wang2024unveiling} introduced a bi-level optimization framework called Graph \underline{C}ontrastive \underline{Lea}rning \underline{R}ecommendation Attack (CLeaR), which is explicitly designed to enable a systematic investigation of the susceptibility of GCL-based recommendation models to targeted promotion attacks. CLeaR provides a new perspective to examine how the spectral smoothness effect induced by graph contrastive learning alters the exposure patterns of individual items. At its core, CLeaR incorporates two main objectives: the dispersion promotion objective, which amplifies spectral smoothness to facilitate embedding dispersion and thus passively creates favorable conditions for targeted promotion; and the rank promotion objective, which actively increases the visibility and exposure of target items to a broad user base.

Nevertheless, recognizing the vulnerability of GCL-based models to such attacks, a fundamental defense dilemma persists: many existing defense strategies risk disrupting the intrinsic spectral smoothness that is central to GCL’s strong recommendation performance. To address this dilemma, we propose a unified two-stage defense framework, named \underline{S}pectral-\underline{I}rregularity \underline{M}itigation(SIM). In the first stage, anomaly detection is conducted by systematically analyzing low-rank spectral irregularities in item embeddings to identify candidate target items that deviate from the expected smooth manifold. In the following adversarial suppression stage, we introduce a mitigation loss that penalizes excessive spectral smoothness of target-item embeddings relative to user representations, thereby dispersing item clusters and reducing their undue prominence in recommendation results. Our contributions are fourfold: 
\begin{itemize}[leftmargin=*]
\item We reveal and theoretically explain a previously unrecognized vulnerability in GCL-based recommendation models, demonstrating their increased susceptibility to targeted promotion attacks and providing comprehensive empirical validation of this phenomenon.
\item We develop CLeaR, a bi-level optimization framework that enables systematic investigation of the vulnerability of GCL-based recommenders to targeted promotion attacks. 
\item We design a spectral smoothness-based defense model to reduce the harm caused by targeted promotion attacks on GCL-based systems.
\item Extensive experiments demonstrate the increased vulnerability of GCL-based recommenders to targeted promotion attacks, and show that our proposed defense method can effectively counteract the risks of malicious item exposure and system manipulation.
\end{itemize}

To facilitate a systematic understanding of our work, the remainder of this paper is organized as follows. Section \ref{sec_Preliminaries} provides the necessary preliminaries, including a formal definition of the graph recommendation task and the targeted promotion attack scenario. Section \ref{sec_Revisiting} revisits the susceptibility of GCL-based recommender systems from a spectral perspective, offering both empirical evidence and theoretical analysis of the observed vulnerabilities. Section \ref{sec_attack} presents our proposed attack method, CLeaR, detailing its design principles and optimization objectives. Section \ref{sec_defense} introduces our defense framework, SIM, which encompasses both anomaly detection and adversarial suppression phases. Section \ref{sec_Experiments} reports extensive experimental results, including performance comparisons, parameter analyses, and ablation studies. Section \ref{sec_related} reviews related literature, and Section \ref{sec_conclusion} concludes the paper with a discussion of future work.

\section{Preliminaries}
\label{sec_Preliminaries}
\subsection{Graph Recommendation Models}
\stitle{Graph Recommendation Task.}
Let \(\mathcal{G} = (\mathcal{U}\cup \mathcal{I}, \mathcal{E})\) be a bipartite interaction graph, where \(\mathcal{U}\) and \(\mathcal{I}\) are the sets of user and item nodes and \(\mathcal{E}\subseteq (\mathcal{U}\times \mathcal{I}\)) represents observed edges. A graph recommendation model seeks to learn low-dimensional node embeddings \(\mathbf{Z} = \{\mathbf{z}_v \mid v \in (\mathcal{U}\cup \mathcal{I})\}\) that encode both structural and semantic information of \(\mathcal{G}\). For any user \(u\in\mathcal{U}\) and item \(i\in\mathcal{I}\), their similarity score is estimated by the inner product \(\mathbf{z}_u^\top \mathbf{z}_i\). A common pairwise ranking objective is Bayesian Personalized Ranking (BPR) \cite{15rendle2012bpr}:
\begin{equation}
  \mathcal{L}_{\mathrm{rec}}
  = -\,\mathbb{E}_{(u,i,j)\sim P_{\mathcal{E}}}\Bigl[\log\sigma\bigl(\mathbf{z}_u^\top \mathbf{z}_i - \mathbf{z}_u^\top \mathbf{z}_j\bigr)\Bigr],
  \label{eq:bpr_loss}
\end{equation}
where \(\sigma(\cdot)\) is the sigmoid function, \(P_{\mathcal{E}}\) samples triples \((u,i,j)\) with \((u,i)\in\mathcal{E}\) and \((u,j)\notin\mathcal{E}\).

\stitle{Graph Contrastive Learning Task.}
Graph Contrastive Learning (GCL) \cite{16chen2020simple} improves embedding robustness by encouraging consistency between representations from different augmentations. The most commonly used contrastive
loss is InfoNCE \cite{37oord2018representation} formulated as follows:
\begin{equation}
\begin{aligned}
\mathcal{L}_{gcl} = -\sum_{p \in (\mathcal{U} \cup \mathcal{I})} \log \frac{\exp\left( {\overline{\textbf{z}}'_p}^T \overline{\textbf{z}}''_p / \tau \right)}{\sum_{n \in \mathcal{(\mathcal{U} \cup \mathcal{I})}} \exp \left( {\overline{\textbf{z}}'_p}^T \overline{\textbf{z}}''_{n}/ \tau \right)},
\end{aligned}
\label{CLloss}
\end{equation}
where \(\tau\) is a temperature hyperparameter and $\overline{\mathbf{z}} = \frac{\mathbf{z}}{|\mathbf{z}|_2}$.

\stitle{Joint Optimization.}
In practice, graph recommendation models optimize a combined objective that balances the BPR loss and the GCL regularizer:
\begin{equation}
  \mathcal{L} = \mathcal{L}_{\mathrm{rec}} \;+\; \omega\,\mathcal{L}_{\mathrm{gcl}},
  \label{eq:joint_loss}
\end{equation}
where \(\omega\) controls the relative strength of the contrastive term.

\subsection{Targeted Promotion  Attacks in Recommendation}
\stitle{Adversarial Goal.}  
In graph recommendation systems, targeted promotion attacks aim to elevate the rankings of specific items \(\mathcal{I}^T\) by maximizing the probability that these target items appear in the top-$K$ recommendation lists of as many users as possible.

\stitle{Adversary Knowledge.}  
We consider the adversary to be under a white-box assumption, where the attacker has full visibility into the recommendation model’s parameters and architecture. It is noteworthy that the adaptation of a white-box attack strategy to a black-box environment often entails the use of a proxy model. This approach has been extensively acknowledged and validated in current research \cite{45wu2021ready, 46zhang2020practical}.

\stitle{Adversary Capability.}  
The attacker can introduce a set of malicious accounts \(\mathcal{U}_M\) and generate a controlled interaction set \(\mathcal{G}_M\). Due to resource limitations, the number of injected profiles and the volume of their interactions are constrained: by default, \(|\mathcal{U}_M|\) is limited to 1\% of the legitimate user population, and each malicious account’s interaction count does not exceed the average interactions per real user. This threat model is consistent with previous studies \cite{05gunes2014shilling,04wang2022gray,13zhang2022pipattack}.

\stitle{Formulation as a Bi-Level Optimization Problem.}
We frame the targeted promotion attack as a bi-level optimization problem, composed of two optimization stages. The inner optimization computes the optimal recommendation model parameters by utilizing authentic user interactions alongside a fixed set of malicious interactions. Conversely, the outer optimization strategically manipulates the interactions between malicious users and items, aiming to increase the presence of target items within user recommendation lists. The bi-level optimization framework is formulated as follows:
\begin{equation}
\begin{aligned}
\max\limits_{\mathcal{G}_{M}}&\mathcal{L}_{attack}(\mathcal{G}, \mathcal{G}_{M}, \mathbf{Z}^{*}_{\mathcal{U}}, \mathbf{Z}^{*}_{\mathcal{I}}), \\
\mathrm{s.t.}, \quad  \mathbf{Z}^{*}_{\mathcal{U}},& \mathbf{Z}^{*}_{\mathcal{I}}=\arg\min\limits_{\mathbf{Z}_{\mathcal{U}}, \mathbf{Z}_{\mathcal{I}}}\mathcal{L}_{rec}(f(\mathcal{G},\mathcal{G}_{M})),
\end{aligned}
\label{bi-level optimization}
\end{equation}
where $f$ denotes the recommendation model function, $\mathbf{Z}^{*}_{\mathcal{U}}$ and $\mathbf{Z}^{*}_{\mathcal{I}}$ represent the optimized user and item embeddings, respectively, and $\mathcal{L}_{attack}$ is the utility function measuring attack efficacy. Initially, we set the malicious interaction set $\mathcal{G}_{M} \in \emptyset$, and then iteratively generate new interactions by alternating between inner and outer optimization phases. Because each interaction variable is binary (0 or 1), we follow the greedy relaxation strategy \cite{50zugner2018adversarial}, temporarily treating interactions as continuous during the optimization. After convergence, we discretize by selecting the highest scoring interactions subject to each malicious user’s budget (their total interaction allowance minus the number of target items) and allocate any remaining capacity exclusively to interactions with the designated targets.

\begin{figure}[t]
    \centering
    \begin{subfigure}[b]{0.98\textwidth}
        \centering
        \includegraphics[width=\textwidth]{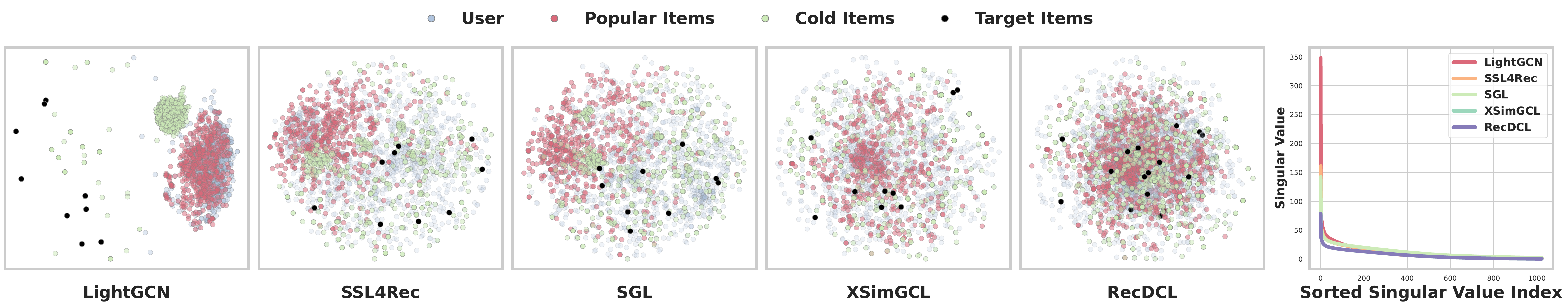}
        \caption{Representation distribution and singular value distributions on the DouBan dataset under RandomAttack.}
    \end{subfigure}    
    \hfill
    \begin{subfigure}[b]{0.98\textwidth}
        \centering
        \includegraphics[width=\textwidth]{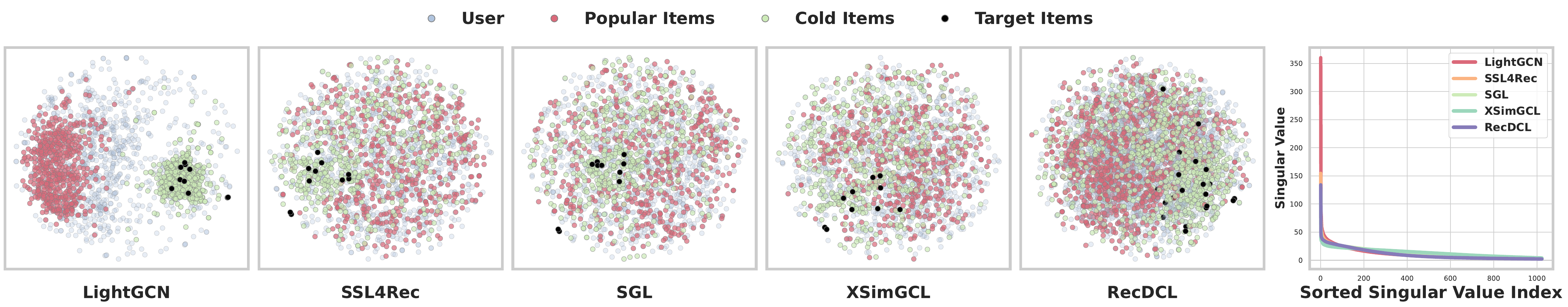}
        \caption{Representation distribution and singular value distributions on the Epinions dataset under RandomAttack.}
    \end{subfigure}
    \hfill
    \begin{subfigure}[b]{0.98\textwidth}
        \centering
        \includegraphics[width=\textwidth]{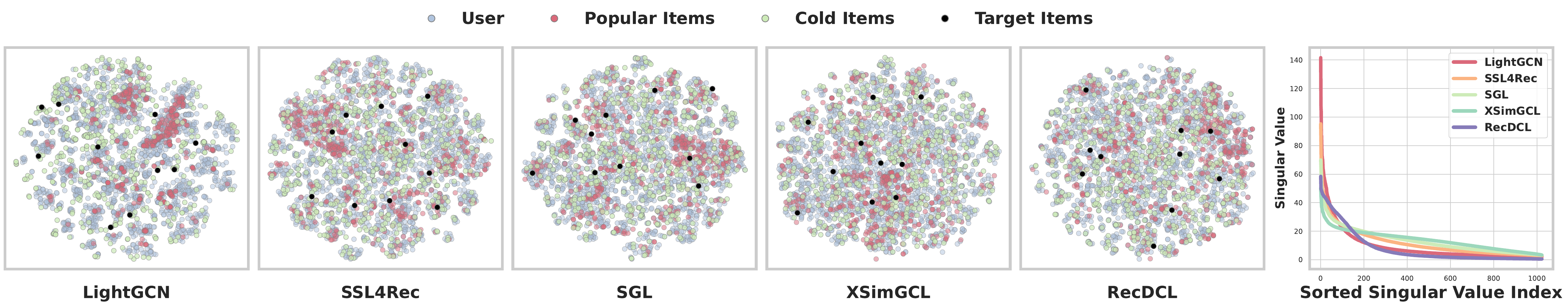}
        \caption{Representation distribution and singular value distributions on the Gowalla dataset under RandomAttack.}
    \end{subfigure}    
    \caption{Representation distribution and singular value distributions on three datasets under RandomAttack.}
    \label{spectralvalue}
\end{figure}

\section{Revisiting Susceptibility of GCL-Based Recommendation from a Spectral Perspective}
\label{sec_Revisiting}
Given the heightened vulnerability of GCL-based recommender systems to targeted promotion attacks, it becomes essential to explore the fundamental causes behind this susceptibility. Traditional evaluation metrics, while informative about overall recommendation performance, are limited in that they overlook the underlying structure and dispersion of learned embeddings. Recent studies~\cite{81chen2024towards,82jing2021understanding,83lin2025recommendation} have highlighted that the spectral properties of the embedding space play a pivotal role in shaping a model’s robustness and its ability to utilize information effectively. Motivated by these insights, we adopt a spectral analysis perspective in this section. Spectral analysis offers a unique lens to investigate how contrastive optimization influences the organization and spread of embeddings. Specifically, the singular value spectrum of the embedding matrix reveals how variance is allocated across dimensions, thereby indicating the model’s efficiency in utilizing its representational capacity. Building on this, we systematically examine the spectral vulnerabilities of GCL-based models through both empirical experiments and theoretical analysis.

\subsection{Empirical Observations}
First, we begin by empirically examining how graph contrastive learning alters the spectral properties of the embedding space, thereby reshaping item representations and influencing item exposure in the presence of targeted promotion attacks. The experimental setup
employed is as follows: To mitigate bias from the power law distribution of item popularity, we sort items by their total interactions, then randomly draw 500 popular items from the top 5\% and 500 cold items from the remaining 80\%. From this pool of cold items, we designate 10 as targets, and we randomly select 500 users. We then retrain both LightGCN and its GCL augmented variant on the resulting poisoned interaction graph and visualize the embedding spaces before and after applying the contrastive learning loss. After training, we extract the learned node embeddings and apply t-SNE \cite{17van2008visualizing} to project them into two dimensions. We also perform singular value decomposition (SVD) \cite{54hoecker1996svd} on the full embedding matrix and sort the singular values in descending order. Fig.~\ref{spectralvalue} presents both the t-SNE projections and the singular-value decay curves for LightGCN without GCL and the GCL-enhanced model on the DouBan~\cite{32Zhao}, Epinions~\cite{58pan2020learning}, and Gowalla~\cite{25yu2022xsimgcl} datasets. 

In our t-SNE projections of the learned embedding spaces, the clustering and dispersion patterns are closely related to the application scenarios and interaction structures of each dataset. Both DouBan (Chinese movie ratings) and Epinions (product reviews) datasets are characterized by centralized user engagement, where users tend to interact heavily with a small number of highly popular items (blockbuster movies or trending products). As a result, these datasets exhibit pronounced clustering: popular items and their most active users form dense clusters at the center of the embedding space, while cold items—those with few interactions—are pushed to the periphery, clearly reflecting the effect of popularity bias. In contrast, the Gowalla dataset, which records user check-ins at physical locations, reflects a scenario where user interests and item exposure are more geographically and behaviorally diverse. Here, although the overall clustering of users and items is less sharp compared to DouBan and Epinions, popular venues (items) and the users who frequently check in still form tightly aggregated groups within the embedding space. Cold locations, visited infrequently, remain relatively isolated from these central clusters. In summary, under LightGCN, all three datasets demonstrate a strong aggregation of popular items and users. This structure makes it particularly difficult for target items, typically cold items with low engagement, to be included in users’ recommendation lists.

When LightGCN is augmented with GCL, the embeddings on all datasets spread more evenly across the two-dimensional map. The node vectors, particularly those of users, fill the entire projection area, and the distinction between popular and cold items narrows substantially. On DouBan and Epinions, this leads to a significant increase in the visibility of cold items, effectively mitigating the popularity bias inherent in the original model. For Gowalla, although the clustering is not as prominent or exclusive as in the other two datasets, it is still apparent that GCL increases the overall dispersion of user embeddings and reduces the separation between popular and cold items. which results in a more uniform distribution. Across all datasets, these results clearly demonstrate that GCL is responsible for the broader dispersion that brings cold-tail target items closer to the core embedding distribution, thereby making them more susceptible to targeted promotion.

We further examine the singular value spectra of the item embedding matrices using singular value decomposition. All datasets show a typical long-tail decay in singular values, but the patterns differ across models. For LightGCN, the singular values drop sharply after the first few, meaning that most embedding variance is concentrated in a small number of leading directions. In contrast, the GCL-augmented models have a more gradual decay of singular values on all datasets. This indicates that GCL reduces the dominance of the largest singular values and spreads variance more evenly across many dimensions, which in turn leads to a more uniform distribution of item and user embeddings in the representation space. However, this more uniform embedding distribution also lowers the barrier for adversaries, as target items naturally become closer to a larger number of users in the embedding space. As a result, it becomes easier for target items to be included in users’ recommendation lists during targeted promotion attacks.


\subsection{Theoretical Analysis}
Beyond empirical findings, we further provide a theoretical perspective to explain why GCL-based models exhibit increased susceptibility to targeted promotion, focusing on the spectral behavior of their embedding spaces. Specifically, we perform a singular value decomposition of the representation matrix $\textbf{Z}$, i.e., $\textbf{Z} = \mathbf{L} \mathbf{\Sigma} \mathbf{R}^T$, where $\mathbf{L} = (\textbf{l}_{1}, ..., \textbf{l}_{d})$ represents $d$-order orthogonal matrix whose columns $\textbf{l}$ are the left singular vectors, and $\mathbf{R} = (\textbf{r}_{1}, ..., \textbf{r}_{N})$ represents $N$-order orthogonal matrix whose columns $\textbf{r}$ are the right singular vectors. $\mathbf{\Sigma} = diag(\mathbf{\sigma}_{1}, ..., \mathbf{\sigma}_{d})$ is the $d \times N$ rectangular diagonal matrix, with non-negative singular values $\sigma$ arranged in descending order. Here, $d$ is the dimensional length of the representation vector, while $N$ is the number of all users and items. Based on this decomposition, we derive Proposition 1, which provides an upper bound for the GCL loss:

\stitle{Proposition 1}. \textit{Give the representations $\textbf{Z}^\prime$ \textit{and} $\textbf{Z}^{\prime\prime}$ which are learned on augmented views} \textit{and the corresponding singular values } $\mathbf{\Sigma}' = diag(\mathbf{\sigma}'_{1}, ..., \mathbf{\sigma}'_{d})$ \textit{and} $\mathbf{\Sigma}'' = diag(\mathbf{\sigma}''_{1}, ..., \mathbf{\sigma}''_{d})$, \textit{an upper bound of the GCL loss is given by:}
\begin{equation}
\begin{aligned}
\mathcal{L}_{gcl} < N \max_{j} \sigma'_{j}  \sigma''_{j} -\sum_{i} \sigma'_{i} \sigma''_{i} + N \log N.
\end{aligned}
\end{equation}

Since minimizing the GCL loss coincides with minimizing this upper bound, Proposition 1 indicates that the minimization of the GCL loss can be reformulated as a dual objective optimization problem: minimizing the first term $\max_{j} \sigma'_{j}  \sigma''_{j}$, which captures the largest singular value product, and maximizing the second term $\bm{\vec{\sigma}}'^{T}\bm{\vec{\sigma}}''$, which corresponds to the sum of pairwise singular value products. This formulation seeks to suppress the strongest singular value correlations while simultaneously strengthening the overall alignment between the singular spectra of $\textbf{Z}'$ and $\textbf{Z}''$. As a result, the optimization naturally promotes a smoother distribution of singular values.

As a consequence of Proposition 1, smoothing the singular‐value spectrum forces more dimensions to carry a meaningful variance, rather than concentrating it on a few top components. In practice, this means that no single direction in $\mathbf{Z}$ dominates the geometry of the embedding space, so node vectors naturally fan out into previously under‐utilized dimensions. Such an isotropic shift explains why, in our t-SNE plots, the GCL‐augmented model yields broadly dispersed clusters instead of the tight, popularity‐driven clusters seen in LightGCN: by punishing the largest singular‐value products and uplifting the smaller ones, GCL promotes a more uniform spread of representations, exactly the global dispersion that exposes cold‐tail and target items more.

We now present the proof of Proposition 1 in detail:

\stitle{Proof of Proposition 1.} For simplification, we first set $\tau=1$ in Eq.~\eqref{CLloss}, and we can get that:
\begin{equation}
\begin{aligned}
\mathcal{L}_{gcl} &= - \sum_{p \in \mathcal{(U\cup I)}} \overline{\mathbf{z}}'^T_p \overline{\mathbf{z}}''_p + \sum_{p \in \mathcal{(U\cup I)}} \log  \sum_{n \in \mathcal{(U\cup I)}} \exp (\overline{\mathbf{z}}'^T_p \overline{\mathbf{z}}''_{n})\\
&< - \sum_{p \in \mathcal{(U\cup I)}} \overline{\mathbf{z}}'^T_p \overline{\mathbf{z}}''_p +  \sum_{p \in \mathcal{(U\cup I)}} \log (N \max_{n \in \mathcal{(U\cup I)}} \exp(\overline{\mathbf{z}}'^T_p \overline{\mathbf{z}}''_{n}))\\
&< - \sum_{p \in \mathcal{(U\cup I)}} \overline{\mathbf{z}}'^T_p \overline{\mathbf{z}}''_p + \sum_{p \in \mathcal{(U\cup I)}}( \log N + \max_{n \in \mathcal{(U\cup I)}} \overline{\mathbf{z}}'^T_n \overline{\mathbf{z}}''_{n}).
\end{aligned}
\end{equation}
In the equation shown, the step from the first to the second line employs the log-sum-exp trick, collapsing all terms of the form $\overline{\mathbf{z}}'^T_p \overline{\mathbf{z}}''_{n}$ to their maximum value and thereby streamlining the computation. In the transition to the third line, motivated by the analyses in \cite{19wang2020understanding, 49wang2022towards}, we assume that the two augmented representations are closely aligned. Under this assumption, the largest similarity between any negative pair $\overline{\mathbf{z}}'^T_p \overline{\mathbf{z}}''_{n}$, falls below the maximum similarity within the augmented pair itself $\overline{\mathbf{z}}'^T_n \overline{\mathbf{z}}''_n$. It follows that the contrastive learning objective can be equivalently expressed as:
\begin{equation}
\begin{aligned}
\mathcal{L}_{gcl} &< - \sum_{p \in \mathcal{(U\cup I)}} \overline{\mathbf{z}}'^T_p \overline{\mathbf{z}}''_p + N \max_{n \in \mathcal{(U\cup I)}} (\overline{\mathbf{z}}'^T_n \overline{\mathbf{z}}''_{n}) + N \log N \\
&= - Tr({\textbf{Z}'}^T \textbf{Z}'') + N \max_{n \in \mathcal{(U\cup I)}} (\overline{\mathbf{z}}'^T_n \overline{\mathbf{z}}''_{n}) + N \log N,
\label{tr}
\end{aligned}
\end{equation}
where $Tr(\cdot)$ means the trace of matrix.

Let $\mathbf{Z}'$ and $\mathbf{Z}''$ be the two augmented versions of the original representation. Assuming their singular vectors differ negligibly, we enforce them to be identical. Consequently, we can write $\textbf{Z}'= \mathbf{L} \mathbf{\Sigma}' \mathbf{R}^T$ and $\textbf{Z}''=\mathbf{L} \mathbf{\Sigma}'' \mathbf{R}^T$, which lead to:
\begin{equation}
\begin{aligned}
{\textbf{Z}'}^T \textbf{Z}'' &= \textbf{R} \Sigma'^{T} \textbf{L}^T \textbf{L} \Sigma'' \textbf{R}^T \\
&=\sigma'_{1}\sigma''_{1} \textbf{r}_{1}{\textbf{r}_{1}}^T + \sigma'_{2}\sigma''_{2} \textbf{r}_{2}{\textbf{r}_{2}}^T + \cdots + \sigma'_d\sigma''_{d} \textbf{r}_{d}{\textbf{r}_{d}}^T.
\end{aligned}
\end{equation}

Then, we plug this equation into Eq.~\eqref{tr}, and due to $\textbf{r}^T_{j} \textbf{r}_{j}=1$, we can get that:
\begin{equation}
\begin{aligned}
\mathcal{L}_{gcl} & < -Tr({\textbf{Z}'}^T \textbf{Z}'') + N \max_{n \in \mathcal{(U\cup I)}} (\overline{\mathbf{z}}'^T_n \overline{\mathbf{z}}''_{n}) + N \log N\\
& = -\sum_{i} \sigma'_{i} \sigma''_{i}  + N \max_{j} \sigma'_{j}  \sigma''_{j} \textbf{r}^T_{j} {\textbf{r}_{j}} + N \log N \\
&= N \max_{j} \sigma'_{j}  \sigma''_{j} -\sum_{i} \sigma'_{i} \sigma''_{i} + N \log N.
\end{aligned}
\end{equation}

With this, Proposition 1 is established. The foregoing analysis demonstrates that the spectral smoothing effect induced by contrastive learning fundamentally amplifies the susceptibility of GCL-based recommender systems to targeted promotion attacks. Building upon this theoretical insight, we proceed in the subsequent section to introduce a targeted attack method that enables systematic exploration and assessment of the vulnerabilities arising from this spectral property in GCL-based recommender systems.

\section{Proposed Attack Method}
\label{sec_attack}
\begin{figure}[t]
\centering
\includegraphics[width=0.7\textwidth]{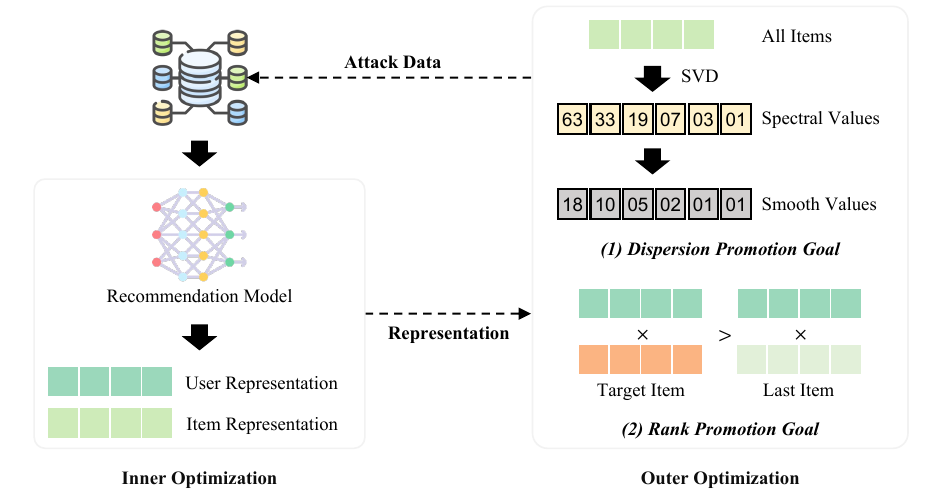}
\caption{The overview of CLeaR.} 
\label{attackmethod}
\end{figure}

Our theoretical investigation and empirical evaluations reveal that the optimization of GCL inherently induces a more uniform spectral value distribution. To further understand how this property affects system security, we propose Graph \underline{C}ontrastive \underline{Lea}rning \underline{R}ecommendation Attack (CLeaR), a targeted promotion attack method that systematically explores the impact of spectral smoothness on recommendation outcomes.  Specifically, CLeaR follows the bi-level optimization framework defined in Equation~\ref{bi-level optimization}. As illustrated in Fig.~\ref{attackmethod}, the optimization process of CLeaR consists of two phases. In the inner optimization phase, optimal user and item embeddings are obtained by training the recommendation model, which serves as the foundation for the subsequent outer optimization phase. The outer optimization is governed by two main objectives, which together form $\mathcal{L}_{attack}$: promoting global dispersion of the embedding space and explicitly increasing the ranking of target items. Through this design, CLeaR reveals how enhanced spectral smoothness can increase the exposure risk of targeted items in GCL-based recommenders.

\subsection{Dispersion Promotion Goal}
The first dispersion promotion goal encourages a more uniform spectral value distribution, thereby ensuring that embeddings are evenly spread across the representation space. This is formalized by aligning the spectral values of the embeddings with a smoother power-law distribution. Embeddings $\textbf{Z}$ undergo singular value decomposition (SVD) to obtain the spectral values ($\bm{\vec{\sigma}}$), and the dispersion objective $\mathcal{L}_D$ can be defined as:
\begin{equation}
\begin{aligned}
\mathcal{L}_{D} = sim (\bm{\vec{\sigma}}, cx^{-\beta}),
\end{aligned}
\label{simdistribution}
\end{equation}
where $sim(\cdot)$ evaluates similarity between distributions, and $cx^{-\beta}$ represents a power-law distribution parameterized by constants $c$ and $\beta$. Reducing $\beta$ smoothens the tail of the spectral distribution.

However, directly applying SVD presents computational inefficiency and numerical instability during gradient backpropagation. To circumvent these limitations, we adopt a spectral approximation technique inspired by rank-1 approximations \cite{56zhang2023spectral,57yu2020toward}. This technique bypasses explicit SVD computation, instead directly deriving approximate embeddings $\mathbf{\hat{Z}}$:
\begin{equation}
\begin{aligned}
\mathbf{\hat{Z}}=\textbf{Z} - \frac{\textbf{Z}\textbf{V}'\textbf{V}'^{T}}{||\textbf{V}'||_{2}^{2}},
\end{aligned}
\end{equation}
where $\textbf{V}'=\textbf{Z}^{T}\textbf{Z}\textbf{V}$ is randomly initialized each iteration. This method indirectly regularizes singular values, compressing large singular values and boosting smaller ones to achieve uniformity.

With $\mathbf{\hat{Z}}$, the iterative SVD and backpropagation are avoided, reformulating Eq.~\eqref{simdistribution} into an $L_{1}$ similarity between original and approximated embeddings:
\begin{equation}
\mathcal{L}_{D} = sim (\mathbf{Z}, \mathbf{\hat{Z}}) 
=- \frac{||\textbf{Z}\textbf{V}'\textbf{V}'^{T}||}{||\textbf{V}'||_{2}^{2}}.
\end{equation}

\subsection{Rank Promotion Goal}
Although embedding dispersion enhances the visibility of target items, it does not explicitly guarantee their prioritization. Thus, the second rank promotion goal focuses on improving the ranking of target items within user recommendation lists. Drawing inspiration from recent targeted attack strategies, we utilize a differentiable CW loss \cite{20rong2022fedrecattack} to explicitly encourage the promotion of target items relative to the lowest-ranked recommended items:
\begin{equation}
\begin{aligned}
\mathcal{L}_{R}=&\sum_{u \in (\mathcal{U})}\sum_{t \in \mathcal{I}^{T} \land \left(u, t\right) \notin \mathcal{G}} g(\mathbf{z}_u^T \mathbf{z}_t - \min_{i \in {\mathcal{I}_u^{\text{rec}} \land i\notin \mathcal{I}^{T}}}\left\{\mathbf{z}_u^T \mathbf{z}_i\right\}),\\
&where \quad g(x)= \begin{cases}x, & x \geq 0, \\ e^x-1, & x<0,\end{cases}
\end{aligned}
\end{equation}
where $\min_{i \in {\mathcal{I}_u^{\text{rec}} \wedge i\notin \mathcal{I}^{T}}}\left\{\mathbf{z}_u^T \mathbf{z}_i\right\}$ indicates the lowest-ranked item recommended to user $u$, and $\mathcal{I}^{rec}_{u}$ denotes items recommended for user $u$. 

Ultimately, we integrate both objectives, defining the total attack loss as:
\begin{equation}
\begin{aligned}
\mathcal{L}_{attack}=\mathcal{L}_{D}+\alpha \mathcal{L}_{R},
\end{aligned}
\end{equation}
where $\alpha$ is the weight for adjusting the balance. 

Based on the aforementioned description, we present the attack algorithm of CLeaR referred to as Algorithm \ref{algorithm1}.

\begin{algorithm}[h]
  \caption{The Process of CLeaR}
  \label{algorithm1}
  \begin{algorithmic}[1]
    \REQUIRE Recommendation model $f$, malicious users $\mathcal{U}_M$, target items $\mathcal{I}^T$, interaction data $\mathcal{G}$
    \STATE Initialize $\mathcal{G}_M \gets \emptyset$
    \WHILE{not converged}
      \STATE \{Inner optimization\}
      \STATE Sample tuples $(u,i,j)$ from $\mathcal{G}\cup\mathcal{G}_M$
      \STATE Compute $L_{\mathrm{rec}}$ to obtain user embeddings and item embeddings
      \STATE \{Outer optimization\}
      \STATE Compute dispersion promotion objective $\mathcal{L}_D$
      \STATE Compute rank promotion objective $\mathcal{L}_R$
      \STATE Minimize $\mathcal{L}_{\mathrm{attack}}$ to obtain the optimal user and item embeddings
      \STATE For each malicious user, greedily select items whose embeddings are closest to the malicious user embedding as interaction items, and update $\mathcal{G}_M$ accordingly
    \ENDWHILE
    \RETURN $\mathcal{G}_M$
  \end{algorithmic}
\end{algorithm}

\section{Proposed Defense Method}
\label{sec_defense}
In the previous sections, we noticed that GCL inherently promotes a smooth spectral distribution within learned embeddings, effectively regularizing embedding dimensions and enhancing representational robustness. However, this intrinsic smoothness can be exploited by attackers through targeted promotion perturbations, artificially boosting the visibility of specific items. To systematically address this vulnerability, we propose a defense method named \underline{S}pectral \underline{I}rregularity \underline{M}itigation (SIM), which preserves the spectral smoothness required for strong recommendation performance while mitigating the harmful effects of targeted promotion attacks.

Our SIM defense strategy leverages the fundamental property that item embeddings—whether align closely with a dominant low-rank subspace formed by the leading singular vectors, although for GCL-based methods, this is relatively mitigated. This alignment arises because most genuine item embeddings share common underlying structural characteristics, allowing them to be effectively captured by a limited set of principal spectral directions and thus yielding low reconstruction errors when projected onto this subspace. In this context, targeted promotion attacks artificially link specific target items to an excessively broad and unnatural set of user embeddings. These manipulated interactions disrupt the typical spectral alignment, forcing targeted items to diverge from the principal subspace. As a result, the reconstruction errors of these target items become substantially larger than those of normal items.

\begin{figure}[t]
\centering
\includegraphics[width=0.8\textwidth]{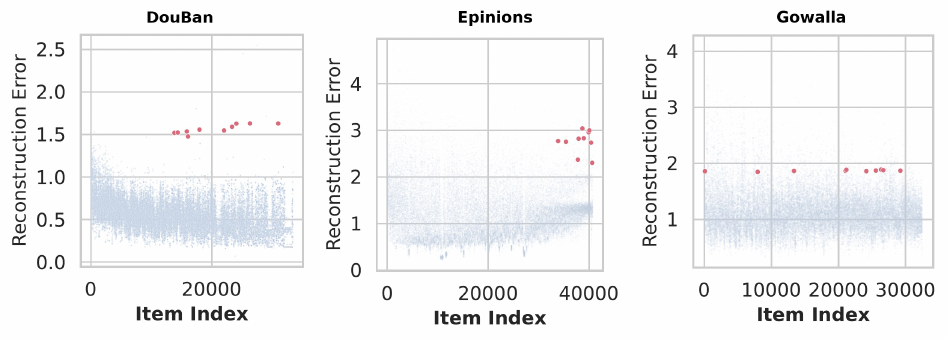}
\caption{Visualization of spectral irregularities across datasets. Red dots represent target items, and gray dots denote normal items. Target items clearly exhibit higher reconstruction errors than normal items.}
\label{visualization}
\end{figure}

However, as shown in Fig.~\ref{spectralvalue}, the principal subspace of non-GCL recommendation models is often overly narrow due to an extremely dominant leading singular value—a phenomenon known as dimensional collapse~\cite{81chen2024towards,82jing2021understanding}. In such cases, even normal items may not align well with the principal subspace, blurring the distinction between normal and abnormal items and making spectral irregularities difficult to detect. In contrast, GCL-based methods reduce the dominance of the leading singular value and expand and stabilize the principal subspace. This richer and more expressive spectral structure enables the vast majority of genuine item embeddings to be effectively aligned and reconstructed with low error. Consequently, items that cannot be captured by this enriched subspace, such as those manipulated by targeted attacks, stand out sharply due to their significantly elevated reconstruction errors. These pronounced spectral deviations thus provide clear and reliable indicators of malicious manipulation, enabling precise detection of abnormal items.

To substantiate the existence and detectability of these spectral irregularities, we conduct an empirical visualization analysis using the XSimGCL model under the CLeaR attack. Specifically, we first apply a rank-$k$ SVD on the learned item embedding matrix, decomposing it as $\mathbf{Z}_{\mathcal{I}} \approx \mathbf{L}^{top-k} \mathbf{\Sigma}^{top-k} \mathbf{R}^{top-k}$. Here, $\mathbf{L}^{top-k}$ and $\mathbf{R}^{top-k}$ represent the top-$k$ left and right singular vectors, respectively, and $\mathbf{\Sigma}^{top-k}$ contains the leading singular values. Each item embedding $\mathbf{z}_i$ is then projected back onto this dominant subspace to obtain its spectral approximation. We quantify the degree of spectral irregularity for each item using the reconstruction error:
\begin{equation}
\epsilon_i = | \mathbf{z}_i - \mathbf{L}^{top-k} \mathbf{\Sigma}^{top-k} \mathbf{R}^{top-k} |_2,
\label{reconstruction}
\end{equation}
where a larger $\epsilon_i$ indicates that $z_i$ cannot be effectively captured by the principal subspace, thus highlighting a potential anomaly. Figure~\ref{visualization} illustrates the reconstruction errors $\epsilon_i$ of items across DouBan, Epinions, and Gowalla datasets. Target items (red dots) exhibit notably higher reconstruction errors compared to normal items (gray dots), highlighting significant spectral deviations induced by targeted attacks. This empirical observation confirms the effectiveness of spectral deviation as an indicator of detecting potential poisoning attacks in GCL-based recommender systems.

\begin{figure}[t]
\centering
\includegraphics[width=0.7\textwidth]{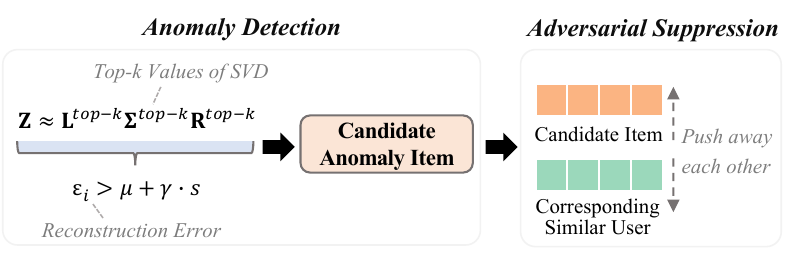}
\caption{The overview of SIM.} 
\label{defense}
\end{figure}

Motivated by these empirical findings, the SIM leverages the contrast between the normal spectral smoothness of genuine items and the irregularities induced by targeted attacks, which systematically operates in two stages: (1) an anomaly detection phase, which systematically identifies candidate target items exhibiting abnormal spectral signatures, and (2) an adversarial suppression phase, which selectively mitigates the influence of these anomalies without compromising the overall spectral integrity of the model. Each phase is detailed below. The main components of our proposed method are illustrated in Fig.~\ref{defense}.

\subsection{Anomaly Detection Phase}
The anomaly detection phase begins by computing the reconstruction error $\epsilon_i$ for each item as defined in Eq.~\eqref{reconstruction}. Since genuine items typically conform to the underlying low-rank manifold, their reconstruction errors cluster within a narrow range. In contrast, items subjected to targeted manipulation deviate substantially. To adaptively and robustly identify such anomalies, we introduce an adaptive statistical thresholding approach, flagging items whose reconstruction error significantly exceeds the normal range:
\begin{equation}
\label{gamma}
\mathcal{T} = \left\{ i \mid \epsilon_i \geq \mu + \gamma \cdot s \right\},
\end{equation}
where $\mu$ and $s$ denote the mean and standard deviation of reconstruction errors across all items, respectively, and $\gamma$ is an adjustable threshold parameter balancing sensitivity and specificity. 
A larger \(\gamma\) yields fewer detected anomalies, while a smaller \(\gamma\) increases sensitivity.

\subsection{Adversarial Suppression Phase}
Upon identifying anomalous items, direct removal is avoided as it may risk discarding legitimate interactions. Instead, a targeted mitigation loss is introduced, which suppresses the harmful influence associated with candidate target items while preserving the spectral smoothness of normal items in GCL, thus maintaining overall model performance. Specifically, the Top-$\text{m}$ users most closely aligned (i.e., with the highest cosine similarity) to each candidate anomalous item are selected, and their excessive embedding similarities are penalized as follows:
\begin{equation}
\mathcal{L}_{mit} = \frac{1}{|\mathcal{T}| \cdot m} \sum_{i \in \mathcal{T}} \sum_{u \in \text{Top-m(i)}} \frac{z_i^\top z_{u}}{\|z_i\| \|z_{u}\|},
\label{mitigation}
\end{equation}
where \(\text{Top-m(i)}\) denotes the top-\(m\) users whose embeddings have the highest cosine similarity with item \(i\), and the total loss during the adversarial suppression phase becomes
\begin{equation}
\mathcal{L}_{total}
= \mathcal{L}_{rec} + \omega\,\mathcal{L}_{gcl} + \lambda_{mit}\mathcal{L}_{mit},
\label{lambda}
\end{equation}
where $\lambda_{mit}$ is a hyperparameter controlling the strength of the mitigation.

It is worth noting that, since SVD is computationally intensive, we adopt an efficient strategy in practice: after each training epoch, anomaly detection is performed to identify candidate targeted items. In the subsequent epoch, the set of candidate targeted items is fixed, and their corresponding top-$m$ most similar users are determined to compute $\mathcal{L}_{mit}$. Through these two stages, SIM ensures robust detection of targeted promotion anomalies and significantly mitigates their impact, effectively maintaining the integrity of recommendation outputs. The complete detection and mitigation procedure is summarized in Algorithm~\ref{algorithm2}.

\begin{algorithm}[h]
  \caption{The process of SIM}
  \label{algorithm2}
  \begin{algorithmic}[1]
    \REQUIRE Item embeddings $\mathbf{Z}_{\mathcal{I}}$, user embeddings $\mathbf{Z}_{\mathcal{U}}$, number of SVD components $k$, threshold parameter $\gamma$, mitigation weight $\lambda_{mit}$, number of top users $m$

    \STATE \COMMENT{Phase 1: Anomaly Detection}
    \STATE Compute rank-$k$ SVD: $\mathbf{Z}_{\mathcal{I}} \approx \mathbf{L}^{top-k} \mathbf{\Sigma}^{top-k} \mathbf{R}^{top-k}$
    \FOR{each item $i$}
        \STATE Compute reconstruction error $\epsilon_i$
    \ENDFOR
    \STATE Compute mean $\mu$ and standard deviation $s$ of $\{\epsilon_i\}$
    \STATE Flag items with $\epsilon_i \geq \mu + \gamma \cdot s$ as anomalous items $\mathcal{T}$

    \STATE \COMMENT{Phase 2: Adversarial Suppression}
    \WHILE{not converged}
      \STATE Sample mini-batch of users and items $(u, i, j)$
      \STATE Compute standard recommendation loss $\mathcal{L}_{rec}$ and contrastive loss $\mathcal{L}_{gcl}$
      \FOR{each $i \in \mathcal{T}$}
        \STATE Find top-$m$ users with highest cosine similarity to $z_i$
      \ENDFOR
      \STATE Compute mitigation loss $\mathcal{L}_{mit}$
      \STATE Update embeddings by minimizing $\mathcal{L}_{total}$
    \ENDWHILE

    \RETURN Robust embeddings $\mathbf{Z}_{\mathcal{U}}, \mathbf{Z}_{\mathcal{I}}$
  \end{algorithmic}
\end{algorithm}

\section{Experiments}
\label{sec_Experiments}
\subsection{Experimental Settings}

\noindent\textbf{Datasets.} We evaluate on four public benchmarks of varying scales: DouBan \cite{32Zhao}, Epinions \cite{58pan2020learning}, and Gowalla \cite{25yu2022xsimgcl}. Statistics are summarized in Table~\ref{datasets}.

\noindent\textbf{Evaluation Protocol.} Each dataset is randomly split into training, validation and test sets in a 7:1:2 ratio. Poisoning perturbations are applied only to the training portion. In our experiments, all metric results are averaged over 5 independent runs, and we report two metrics:

\textbf{Recall@50}, measuring recommendation quality:
    \begin{equation}
      \mathrm{Recall}@K
      =\frac{1}{|\mathcal{U}|}\sum_{u\in\mathcal{U}}
      \frac{\bigl|\mathrm{Top}_{K}(u)\cap P_u\bigr|}{|P_u|}\,,
    \end{equation}
    where $P_u$ is the set of held-out items for user $u$ and $\mathrm{Top}_{K}(u)$ are the top-50 recommendations for user $u$.

\textbf{Hit Ratio@50}, measuring attack success on target items:
\begin{equation}
  \mathrm{HR}@K
  =\frac{1}{|\mathcal{U} \setminus \mathcal{U}_M|}\sum_{u\in\mathcal{U} \setminus \mathcal{U}_M}
  \mathbb{I}\bigl(\mathcal{I}^T\in\mathrm{Top}_{K}(u)\bigr)\,.
\end{equation}

\noindent\textbf{Recommendation Methods.} We use LightGCN \cite{07he2020lightgcn} as the backbone and compare four recent GCL-based variants:
\begin{itemize}
  \item SSL4Rec \cite{55yao2021self}:  incorporates an item-level contrastive learning objective to enhance the discrimination of item embeddings. By contrasting item representations across different augmentations, SSL4Rec aims to improve robustness against noisy or missing interactions.
  \item SGL \cite{08wu2021self}: applies random node and edge dropout strategies to generate diverse graph augmentations for contrastive learning. This process encourages the model to learn invariant representations under different structural perturbations.
  \item XSimGCL \cite{25yu2022xsimgcl}: streamlines the graph augmentation process and emphasizes uniformity in the representation space. It adopts a simple yet effective perturbation at the embedding level, significantly reducing computational overhead.
  \item RecDCL \cite{72zhang2024recdcl}: combines both batch-wise and feature-wise contrastive losses to simultaneously capture instance-level and feature-level similarities. This dual contrastive strategy further enriches the learned representations for recommendation.
\end{itemize}

\noindent\textbf{Attack Baselines.} We benchmark against five representative targeted promotion attack strategies:
\begin{itemize}
  \item RandomAttack \cite{21li2016data}: simulates shilling by interacting with both target items and randomly selected items. It serves as a simple baseline to evaluate the effectiveness of more advanced attack strategies.
  \item AUSH \cite{22lin2020attacking}: leverages Generative Adversarial Networks (GANs) to generate synthetic malicious user profiles. The generated profiles are designed to mimic real user behaviors while promoting specific target items.
  \item GOAT \cite{45wu2021ready}: integrates graph convolutional networks into the GAN framework to produce smoother and more realistic fake ratings. This method is effective in evading detection while maximizing target item exposure.
  \item AutoAttack \cite{73guo2023targeted}: automatically learns optimal feature and edge perturbations to construct malicious interactions. It focuses on influencing specific user groups, making the attack highly targeted and adaptive.
  \item UBA \cite{79wang2024uplift}: assigns individualized costs to crafting fake interactions for each malicious user. By optimizing resource allocation, UBA increases attack efficiency and impact on target items. 
\end{itemize}

For our proposed CLeaR,  it introduces a spectral-focused objective in the attack design, explicitly manipulating the representation spectra. This approach strategically amplifies the exposure of target items by shaping the spectral properties of embeddings.

\noindent\textbf{Defense Baselines.} To evaluate the effectiveness of our defense strategy, we compare SIM with the following baseline methods.
\begin{itemize}
  \item GraphRfi \cite{65zhang2020gcn}:  GraphRfi is a GCN-based framework that jointly performs recommendation and fraudster detection. It dynamically adjusts user influence based on authenticity assessments, effectively filtering out malicious behaviors during recommendation.
  \item BOD \cite{71wang2023efficient}: formulates the denoising task as a bi-level optimization problem. It uses feedback from an inner denoising objective to reweight samples in the outer optimization, aiming to improve robustness against noisy data.
  \item DCF \cite{74he2024double}: introduces a two-stage correction mechanism for iterative sample relabeling. It addresses both observed and unobserved interactions, mitigating the negative impact of mislabeled or adversarial data on recommendation performance.
\end{itemize}

\noindent\textbf{Experimental Details.} We implement all experiments in PyTorch. Unless otherwise specified, we use the recommended parameter settings for all models: batch size is set to 512, learning rate to 0.001, embedding size to 128, and the number of LightGCN layers to 3. For each dataset, we randomly select 10 target items from the set of unpopular items, where unpopular items are defined as the bottom 80\% of items ranked by interaction count. Specifically, the chosen target items are: \{`14708', `31611', `6466', `25014', `17702', `2462', `28290', `26879', `17876', `29052'\} for DouBan, \{`72446', `12058', `19138', `12368', `12754', `4931', `1619', `4423', `43647', `1564'\} for Epinions, and \{`28175', `10238', `30585', `22172', `16495', `26443', `117', `24095', `17285', `5776'\} for Gowalla.

It is worth noting that the experiments in this paper primarily focus on evaluating the defense method. For a comprehensive analysis and additional results regarding the attack method, readers may refer to our previous study~\cite{80wang2024unveiling}.
\begin{table}[t]
  \centering
  \caption{Dataset statistics.}
  \label{datasets}
  \begin{tabular}{@{}ccccc@{}}
    \toprule
    \textbf{Dataset} & \textbf{\#Users} & \textbf{\#Items} & \textbf{\#Interactions} & \textbf{Density} \\
    \midrule
    DouBan   & 2,831   & 36,821   & 805,611   & 0.772\% \\
    Epinions & 46,846  & 40,706   & 305,249   & 0.016\% \\
    Gowalla & 18,737  & 32,495   & 566,791   & 0.093\% \\
    \bottomrule
  \end{tabular}
\end{table}

\subsection{Attack and Defense Performance Analysis}
We first compare the performance of CLeaR with existing targeted promotion attack methods and demonstrate the effectiveness of SIM as a defense in recommendation across three different datasets. The results are presented in Table~\ref{attackcomparison}, in which all H@50 values should be multiplied by $10^{-2}$. For the results of SIM, we report the mitigation outcomes achieved under the constraint that overall recommendation performance is not degraded. According to Table~\ref{attackcomparison}, several observations can be made:
\begin{itemize}
\item Without the SIM defense, CLeaR significantly increases the likelihood that target items appear in the recommendation lists generated by GCL-based models, thereby exposing a pronounced vulnerability to targeted promotion attacks. This result highlights that explicitly modeling and manipulating the representation spectra, an aspect overlooked by existing methods, can reveal the security weaknesses inherent in GCL-based recommenders.
\item Additionally, SIM consistently and effectively mitigates the impact of all evaluated attacks across all three datasets, most notably on the DouBan and Gowalla datasets, where it reduces HitRatio by at least an order of magnitude while maintaining recommendation performance. We attribute this improvement to SIM’s precise identification of target items and its suppression of the similarity between potential target items and highly similar users, which also enhances the diversity of learned representations. 
\item It is worth noting that on the Epinions dataset, while SIM still reduces attack effectiveness and maintains recommendation performance, the improvements are less pronounced compared to the other two datasets. This can likely be attributed to the high sparsity of the Epinions data. In highly sparse datasets, the separation of items from their most similar users may inadvertently disrupt genuine user-item pairs that are already rare, thus limiting the effectiveness of the mitigation. Specifically, when the data are sparse, each user and item have far fewer interactions, and forcibly reducing the similarity between target items and their top users may break legitimate associations that are crucial for learning accurate representations.
\end{itemize}
\begin{summarybox}\textbf{Summary}:
CLeaR demonstrates that exploiting spectral smoothing in GCL-based models can substantially elevate the success rate of targeted promotion attacks.  SIM defense, in turn, effectively and consistently reduces the impact of attacks while maintaining overall recommendation quality. On dense datasets, SIM lowers the HitRatio of target items by an order of magnitude, whereas on sparse datasets, its improvements are more moderate due to the need to preserve rare but genuine associations.
\end{summarybox}

\begin{table*}[t]
\caption{Performance comparison of different recommendation methods with and without SIM under various attack strategies across three datasets, using Recall@50 (R@50) and HitRatio@50 (H@50) as evaluation metrics.}
\label{attackcomparison}
\centering
\resizebox{1\textwidth}{!}{
\begin{tabular}{@{}c|c|cc|cc|cc|cc|cc|cc@{}}
\toprule
\multirow{2}{*}{Dataset} &
  \multirow{2}{*}{Model} &
  \multicolumn{2}{c|}{RandomAttack} &
  \multicolumn{2}{c|}{AUSH} &
  \multicolumn{2}{c|}{GOAT} &
  \multicolumn{2}{c|}{AutoAttack} &
  \multicolumn{2}{c|}{UBA} &
  \multicolumn{2}{c}{CLeaR} \\ \cmidrule(l){3-14} 
 &
   &
  \multicolumn{1}{c|}{R@50} &
  H@50 &
  \multicolumn{1}{c|}{R@50} &
  H@50 &
  \multicolumn{1}{c|}{R@50} &
  H@50 &
  \multicolumn{1}{c|}{R@50} &
  H@50 &
  \multicolumn{1}{c|}{R@50} &
  H@50 &
  \multicolumn{1}{c|}{R@50} &
  H@50 \\ \midrule
\multirow{8}{*}{DouBan} 
  & SSL4Rec      
      & 0.1456 & 0.1538 
      & 0.1509 & 0.1719 
      & 0.1492 & 0.1695 
      & 0.1465 & 0.1417 
      & 0.1460 & 0.1692 
      & 0.1478 & 0.1764 \\
  & \cellcolor{gray!15}SSL4Rec-SIM  
      & \cellcolor{gray!15}\textbf{0.1491}$\uparrow$ & \cellcolor{gray!15}\textbf{0.0000}$\downarrow$
      & \cellcolor{gray!15}\textbf{0.1540}$\uparrow$ & \cellcolor{gray!15}\textbf{0.0284}$\downarrow$
      & \cellcolor{gray!15}\textbf{0.1531}$\uparrow$ & \cellcolor{gray!15}\textbf{0.0000}$\downarrow$
      & \cellcolor{gray!15}\textbf{0.1558}$\uparrow$ & \cellcolor{gray!15}\textbf{0.0000}$\downarrow$
      & \cellcolor{gray!15}\textbf{0.1482}$\uparrow$ & \cellcolor{gray!15}\textbf{0.0000}$\downarrow$
      & \cellcolor{gray!15}\textbf{0.1491}$\uparrow$ & \cellcolor{gray!15}\textbf{0.0000}$\downarrow$ \\
  & SGL          
      & 0.1478 & 0.2305
      & 0.1484 & 0.2609 
      & 0.1474 & 0.2750 
      & 0.1468 & 0.2631 
      & 0.1453 & 0.2563 
      & 0.1328 & 0.3007 \\
  & \cellcolor{gray!15}SGL-SIM      
      & \cellcolor{gray!15}\textbf{0.1555}$\uparrow$ & \cellcolor{gray!15}\textbf{0.0248}$\downarrow$
      & \cellcolor{gray!15}\textbf{0.1553}$\uparrow$ & \cellcolor{gray!15}\textbf{0.0000}$\downarrow$
      & \cellcolor{gray!15}\textbf{0.1482}$\uparrow$ & \cellcolor{gray!15}\textbf{0.0822}$\downarrow$
      & \cellcolor{gray!15}\textbf{0.1504}$\uparrow$ & \cellcolor{gray!15}\textbf{0.0000}$\downarrow$
      & \cellcolor{gray!15}\textbf{0.1504}$\uparrow$ & \cellcolor{gray!15}\textbf{0.0000}$\downarrow$
      & \cellcolor{gray!15}\textbf{0.1464}$\uparrow$ & \cellcolor{gray!15}\textbf{0.0000}$\downarrow$ \\
  & XSimGCL      
      & 0.1473 & 0.2598 
      & 0.1488 & 0.2785 
      & 0.1443 & 0.2457 
      & 0.1423 & 0.2934 
      & 0.1453 & 0.2786 
      & 0.1468 & 0.3589 \\
  & \cellcolor{gray!15}XSimGCL-SIM  
      & \cellcolor{gray!15}\textbf{0.1518}$\uparrow$ & \cellcolor{gray!15}\textbf{0.0000}$\downarrow$
      & \cellcolor{gray!15}\textbf{0.1522}$\uparrow$ & \cellcolor{gray!15}\textbf{0.0348}$\downarrow$
      & \cellcolor{gray!15}\textbf{0.1511}$\uparrow$ & \cellcolor{gray!15}\textbf{0.0000}$\downarrow$
      & \cellcolor{gray!15}\textbf{0.1527}$\uparrow$ & \cellcolor{gray!15}\textbf{0.0000}$\downarrow$
      & \cellcolor{gray!15}\textbf{0.1519}$\uparrow$ & \cellcolor{gray!15}\textbf{0.0000}$\downarrow$
      & \cellcolor{gray!15}\textbf{0.1489}$\uparrow$ & \cellcolor{gray!15}\textbf{0.0000}$\downarrow$ \\
  & RecDCL       
      & 0.1405 & 0.2375 
      & 0.1417 & 0.2628 
      & 0.1422 & 0.2671 
      & 0.1427 & 0.2546 
      & 0.1423 & 0.2473 
      & 0.1453 & 0.2892 \\
  & \cellcolor{gray!15}RecDCL-SIM   
      & \cellcolor{gray!15}\textbf{0.1424}$\uparrow$ & \cellcolor{gray!15}\textbf{0.0127}$\downarrow$
      & \cellcolor{gray!15}\textbf{0.1446}$\uparrow$ & \cellcolor{gray!15}\textbf{0.0324}$\downarrow$
      & \cellcolor{gray!15}\textbf{0.1435}$\uparrow$ & \cellcolor{gray!15}\textbf{0.0213}$\downarrow$
      & \cellcolor{gray!15}\textbf{0.1441}$\uparrow$ & \cellcolor{gray!15}\textbf{0.0182}$\downarrow$
      & \cellcolor{gray!15}\textbf{0.1433}$\uparrow$ & \cellcolor{gray!15}\textbf{0.0139}$\downarrow$
      & \cellcolor{gray!15}\textbf{0.1466}$\uparrow$ & \cellcolor{gray!15}\textbf{0.0174}$\downarrow$ \\ \midrule
\multirow{8}{*}{Epinions}
  & SSL4Rec
    & 0.2234 & 1.6358
    & 0.2251 & 1.7837
    & 0.2225 & 1.8851
    & 0.2234 & 1.5872
    & 0.2214 & 1.6783
    & 0.2233 & 2.2386 \\
  & \cellcolor{gray!15}SSL4Rec-SIM
    & \cellcolor{gray!15}\textbf{0.2325}$\uparrow$ & \cellcolor{gray!15}\textbf{1.4631}$\downarrow$
    & \cellcolor{gray!15}\textbf{0.2285}$\uparrow$ & \cellcolor{gray!15}\textbf{1.4917}$\downarrow$
    & \cellcolor{gray!15}\textbf{0.2271}$\uparrow$ & \cellcolor{gray!15}\textbf{1.6389}$\downarrow$
    & \cellcolor{gray!15}\textbf{0.2244}$\uparrow$ & \cellcolor{gray!15}\textbf{1.2398}$\downarrow$
    & \cellcolor{gray!15}\textbf{0.2222}$\uparrow$ & \cellcolor{gray!15}\textbf{0.7793}$\downarrow$
    & \cellcolor{gray!15}\textbf{0.2260}$\uparrow$ & \cellcolor{gray!15}\textbf{2.0793}$\downarrow$ \\
  & SGL
    & 0.2403 & 1.0586
    & 0.2433 & 2.0425
    & 0.2441 & 2.1964
    & 0.2407 & 1.8224
    & 0.2397 & 0.9214
    & 0.2400 & 2.4637 \\
  & \cellcolor{gray!15}SGL-SIM
    & \cellcolor{gray!15}\textbf{0.2414}$\uparrow$ & \cellcolor{gray!15}\textbf{0.8816}$\downarrow$
    & \cellcolor{gray!15}\textbf{0.2438}$\uparrow$ & \cellcolor{gray!15}\textbf{1.6278}$\downarrow$
    & \cellcolor{gray!15}\textbf{0.2466}$\uparrow$ & \cellcolor{gray!15}\textbf{1.4913}$\downarrow$
    & \cellcolor{gray!15}\textbf{0.2424}$\uparrow$ & \cellcolor{gray!15}\textbf{1.5981}$\downarrow$
    & \cellcolor{gray!15}\textbf{0.2398}$\uparrow$ & \cellcolor{gray!15}\textbf{0.8915}$\downarrow$
    & \cellcolor{gray!15}\textbf{0.2430}$\uparrow$ & \cellcolor{gray!15}\textbf{2.1619}$\downarrow$ \\
  & XSimGCL
    & 0.2372 & 1.3361
    & 0.2403 & 0.4693
    & 0.2402 & 1.0457
    & 0.2312 & 1.1289
    & 0.2397 & 1.2384
    & 0.2392 & 1.6395 \\
  & \cellcolor{gray!15}XSimGCL-SIM
    & \cellcolor{gray!15}\textbf{0.2386}$\uparrow$ & \cellcolor{gray!15}\textbf{1.0183}$\downarrow$
    & \cellcolor{gray!15}\textbf{0.2418}$\uparrow$ & \cellcolor{gray!15}\textbf{0.3538}$\downarrow$
    & \cellcolor{gray!15}\textbf{0.2411}$\uparrow$ & \cellcolor{gray!15}\textbf{0.9317}$\downarrow$
    & \cellcolor{gray!15}\textbf{0.2346} & \cellcolor{gray!15}\textbf{0.8895}$\downarrow$
    & \cellcolor{gray!15}\textbf{0.2407} & \cellcolor{gray!15}\textbf{1.0992}$\downarrow$
    & \cellcolor{gray!15}\textbf{0.2410}$\uparrow$ & \cellcolor{gray!15}\textbf{1.4395}$\downarrow$ \\
  & RecDCL
    & 0.2313 & 1.1538
    & 0.2322 & 1.5482
    & 0.2356 & 1.2937
    & 0.2328 & 1.4538
    & 0.2311 & 0.8572
    & 0.2318 & 1.7539 \\
  & \cellcolor{gray!15}RecDCL-SIM
    & \cellcolor{gray!15}\textbf{0.2364}$\uparrow$ & \cellcolor{gray!15}\textbf{0.6675}$\downarrow$
    & \cellcolor{gray!15}\textbf{0.2385}$\uparrow$ & \cellcolor{gray!15}\textbf{1.2273}$\downarrow$
    & \cellcolor{gray!15}\textbf{0.2368}$\uparrow$ & \cellcolor{gray!15}\textbf{1.1956}$\downarrow$
    & \cellcolor{gray!15}\textbf{0.2341}$\uparrow$ & \cellcolor{gray!15}\textbf{1.0671}$\downarrow$
    & \cellcolor{gray!15}\textbf{0.2328}$\uparrow$ & \cellcolor{gray!15}\textbf{0.7761}$\downarrow$
    & \cellcolor{gray!15}\textbf{0.2348}$\uparrow$ & \cellcolor{gray!15}\textbf{1.3627}$\downarrow$ \\
 \midrule
\multirow{8}{*}{Gowalla}
  & SSL4Rec
    & 0.1831 & 0.1473
    & 0.1844 & 0.1518
    & 0.1852 & 0.1272
    & 0.1822 & 0.1342
    & 0.1840 & 0.1137
    & 0.1844 & 0.1637 \\
  & \cellcolor{gray!15}SSL4Rec-SIM
    & \cellcolor{gray!15}\textbf{0.1841}$\uparrow$ & \cellcolor{gray!15}\textbf{0.0469}$\downarrow$
    & \cellcolor{gray!15}\textbf{0.1851}$\uparrow$ & \cellcolor{gray!15}\textbf{0.0386}$\downarrow$
    & \cellcolor{gray!15}\textbf{0.1866}$\uparrow$ & \cellcolor{gray!15}\textbf{0.0673}$\downarrow$
    & \cellcolor{gray!15}\textbf{0.1856}$\uparrow$ & \cellcolor{gray!15}\textbf{0.0471}$\downarrow$
    & \cellcolor{gray!15}\textbf{0.1852}$\uparrow$ & \cellcolor{gray!15}\textbf{0.0357}$\downarrow$
    & \cellcolor{gray!15}\textbf{0.1882}$\uparrow$ & \cellcolor{gray!15}\textbf{0.0852}$\downarrow$ \\
  & SGL
    & 0.1876 & 0.1023
    & 0.1823 & 0.1375
    & 0.1836 & 0.1174
    & 0.1844 & 0.1385
    & 0.1855 & 0.1498
    & 0.1867 & 0.1894 \\
  & \cellcolor{gray!15}SGL-SIM
    & \cellcolor{gray!15}\textbf{0.1889}$\uparrow$ & \cellcolor{gray!15}\textbf{0.0918}$\downarrow$
    & \cellcolor{gray!15}\textbf{0.1877}$\uparrow$ & \cellcolor{gray!15}\textbf{0.0824}$\downarrow$
    & \cellcolor{gray!15}\textbf{0.1856}$\uparrow$ & \cellcolor{gray!15}\textbf{0.0691}$\downarrow$
    & \cellcolor{gray!15}\textbf{0.1871}$\uparrow$ & \cellcolor{gray!15}\textbf{0.0473}$\downarrow$
    & \cellcolor{gray!15}\textbf{0.1891}$\uparrow$ & \cellcolor{gray!15}\textbf{0.0327}$\downarrow$
    & \cellcolor{gray!15}\textbf{0.1888}$\uparrow$ & \cellcolor{gray!15}\textbf{0.0271}$\downarrow$ \\
  & XSimGCL
    & 0.1888 & 0.1126
    & 0.1895 & 0.0983
    & 0.1889 & 0.1121
    & 0.1868 & 0.1239
    & 0.1866 & 0.1412
    & 0.1884 & 0.1469 \\
  & \cellcolor{gray!15}XSimGCL-SIM
    & \cellcolor{gray!15}\textbf{0.1913}$\uparrow$ & \cellcolor{gray!15}\textbf{0.0237}$\downarrow$
    & \cellcolor{gray!15}\textbf{0.1917}$\uparrow$ & \cellcolor{gray!15}\textbf{0.0193}$\downarrow$
    & \cellcolor{gray!15}\textbf{0.1916}$\uparrow$ & \cellcolor{gray!15}\textbf{0.0354}$\downarrow$
    & \cellcolor{gray!15}\textbf{0.1933}$\uparrow$ & \cellcolor{gray!15}\textbf{0.0598}$\downarrow$
    & \cellcolor{gray!15}\textbf{0.1889}$\uparrow$ & \cellcolor{gray!15}\textbf{0.0385}$\downarrow$
    & \cellcolor{gray!15}\textbf{0.1937}$\uparrow$ & \cellcolor{gray!15}\textbf{0.0296}$\downarrow$ \\
  & RecDCL
    & 0.1823 & 0.1329
    & 0.1834 & 0.1048
    & 0.1811 & 0.1193
    & 0.1803 & 0.1346
    & 0.1860 & 0.1231
    & 0.1826 & 0.1672 \\
  & \cellcolor{gray!15}RecDCL-SIM
    & \cellcolor{gray!15}\textbf{0.1853}$\uparrow$ & \cellcolor{gray!15}\textbf{0.0278}$\downarrow$
    & \cellcolor{gray!15}\textbf{0.1846}$\uparrow$ & \cellcolor{gray!15}\textbf{0.0124}$\downarrow$
    & \cellcolor{gray!15}\textbf{0.1833}$\uparrow$ & \cellcolor{gray!15}\textbf{0.0271}$\downarrow$
    & \cellcolor{gray!15}\textbf{0.1847}$\uparrow$ & \cellcolor{gray!15}\textbf{0.0195}$\downarrow$
    & \cellcolor{gray!15}\textbf{0.1888}$\uparrow$ & \cellcolor{gray!15}\textbf{0.0243}$\downarrow$
    & \cellcolor{gray!15}\textbf{0.1856}$\uparrow$ & \cellcolor{gray!15}\textbf{0.0176}$\downarrow$ \\
 \bottomrule
\end{tabular}
}
\end{table*}

\subsection{Defense Performance Comparison}
This section evaluates the effectiveness of various defense methods in mitigating targeted promotion attacks on the XSimGCL backbone. We benchmark our proposed SIM framework against several representative baselines, including GraphRfi, BOD, and DCF, as reported in Table~\ref{defensecomparison}. Note that all H@50 values in Table~\ref{defensecomparison} should be multiplied by $10^{-2}$ for clarity. Among these methods, GraphRfi shares a similar logic with ours in that it first detects anomalies and then enhances the recommendation model based on the detection results. In contrast, BOD and DCF consider not only malicious users but also noise arising from unintentional user behaviors. From Table~\ref{defensecomparison}, we can get the following conclusions: 
\begin{itemize}
    \item Experimental results show that SIM consistently achieves the lowest H@50 across all attack scenarios, while maintaining competitive R@50 on the DouBan dataset. Apart from SIM, GraphRfi is the most effective in attack mitigation, as evidenced by the second-best H@50 values; however, its R@50 is the lowest among all defense methods. This suggests that while GraphRfi’s approach of detecting and filtering malicious users can suppress attacks, it does not improve recommendation performance. 
    \item On the other hand, BOD and DCF achieve relatively high R@50 scores, but because they focus on general denoising rather than targeted attack detection, they only partially mitigate attacks, leading to limited improvements in H@50. These results demonstrate that the item-level detection and suppression strategy employed by SIM is more effective for defending against targeted promotion, as it provides both robust attack mitigation and strong recommendation quality. We observe similar patterns and conclusions on the other two datasets, further confirming the generality of these findings.
\end{itemize}

\begin{summarybox}\textbf{Summary}:
SIM consistently delivers the strongest attack mitigation and preserves high recommendation quality. GraphRfi effectively suppresses attacks but sacrifices accuracy. BOD and DCF maintain recommendation performance but offer only limited protection.
\end{summarybox}

\begin{table*}[t]
\caption{Performance comparison of different defense methods across three datasets, using Recall@50 (R@50) and HitRatio@50 (H@50) as the metrics. For every attack case, the best results are in bold, and the second-best are underlined.}
\label{defensecomparison}
\centering
\resizebox{1\textwidth}{!}{
\begin{tabular}{@{}c|c|cc|cc|cc|cc|cc|cc@{}}
\toprule
\multirow{2}{*}{Dataset} &
  \multirow{2}{*}{Model} &
  \multicolumn{2}{c|}{RandomAttack} &
  \multicolumn{2}{c|}{AUSH} &
  \multicolumn{2}{c|}{GOAT} &
  \multicolumn{2}{c|}{AutoAttack} &
  \multicolumn{2}{c|}{UBA} &
  \multicolumn{2}{c}{CLeaR} \\ \cmidrule(l){3-14} 
 & & 
  \multicolumn{1}{c|}{R@50} & H@50 &
  \multicolumn{1}{c|}{R@50} & H@50 &
  \multicolumn{1}{c|}{R@50} & H@50 &
  \multicolumn{1}{c|}{R@50} & H@50 &
  \multicolumn{1}{c|}{R@50} & H@50 &
  \multicolumn{1}{c|}{R@50} & H@50 \\ \midrule

\multirow{5}{*}{{DouBan}} 
& Normal        & 0.1473 & 0.2598 
      & 0.1488 & 0.2785 
      & 0.1443 & 0.2457 
      & 0.1423 & 0.2934 
      & 0.1453 & 0.2786 
      & 0.1468 & 0.3589 \\
& GraphRfi & 0.1486 & \underline{0.0862} & 0.1496 & \underline{0.0336} & 0.1452 & \underline{0.0415} & 0.1437 & \underline{0.0332} & 0.1483 & \underline{0.0233} & 0.1448 & \underline{0.0318} \\
& BOD      & \textbf{0.1536} & 0.2248 & \underline{0.1512} & 0.2480 & \underline{0.1474} & 0.2237 & 0.1463 & 0.2445 & \underline{0.1527} & 0.2368 & 0.1439 & 0.2804 \\
& DCF      & 0.1522 & 0.1914 & 0.1504 & 0.1982 & {0.1464} & 0.2145 & \underline{0.1515} & 0.2134 & \textbf{0.1544} & 0.1932 & \underline{0.1458} & 0.2206 \\ \cmidrule(lr){2-14}
& SIM      & \underline{0.1528} & \textbf{0.0000} & \textbf{0.1522} & \textbf{0.0348} & \textbf{0.1511} & \textbf{0.0000} & \textbf{0.1527} & \textbf{0.0000} & 0.1519 & \textbf{0.0000} & \textbf{0.1489} & \textbf{0.0000} \\ \midrule

\multirow{5}{*}{{Epinions}}
& Normal   & 0.2372 & 1.3361
    & 0.2403 & 0.4693
    & 0.2402 & 1.0457
    & 0.2312 & 1.1289
    & 0.2397 & 1.2384
    & 0.2392 & 1.6395 \\
& GraphRfi & 0.2383 & \underline{1.1236} & 0.2413 & \underline{0.3869} & 0.2406 & \underline{0.9989} & 0.2337 & \underline{0.9384} & 0.2402 & \underline{1.1395} & 0.2399 & \underline{1.4793} \\
& BOD & \textbf{0.2414} & 1.3116 & \textbf{0.2438} & 0.4627 & \textbf{0.2424} & 1.0111 & \textbf{0.2368} & 1.1213 & \textbf{0.2437} & 1.1619 & \textbf{0.2445} & 1.6194 \\
& DCF & \underline{0.2411} & 1.2138 & \underline{0.2422} & 0.4482 & \underline{0.2418} & 1.0372 & {0.2328} & 1.0761 & \underline{0.2412} & 1.2139 & \underline{0.2438} & 1.5827 \\ \cmidrule(lr){2-14}
& SIM      & 0.2386 & \textbf{1.0183} & 0.2418 & \textbf{0.3538} & 0.2411 & \textbf{0.9317} & \underline{0.2346} & \textbf{0.8895} & 0.2407 & \textbf{1.0992} & 0.2410 & \textbf{1.4395} \\ \midrule

\multirow{5}{*}{{Gowalla}}
& Normal       & 0.1888 & 0.1126
    & 0.1895 & 0.0983
    & 0.1889 & 0.1121
    & 0.1866 & 0.1239
    & 0.1866 & 0.1412
    & 0.1884 & 0.1469 \\
& GraphRfi & 0.1889 & \underline{0.0415} & 0.1899 & \underline{0.0282} & 0.1895 & \underline{0.0388} & 0.1889 & \underline{0.0835} & 0.1877 & \underline{0.0410} & 0.1899 & \underline{0.0495} \\
& BOD      & \underline{0.1902} & 0.0904 & \textbf{0.1921} & 0.0625 & \underline{0.1901} & 0.1074 & \textbf{0.1936} & 0.1001 & \textbf{0.1914} & 0.1215 & {0.1902} & 0.1255 \\
& DCF      & 0.1867 & 0.0812 & {0.1910} & 0.0718 & 0.1882 & 0.0917 & {0.1915} & 0.1022 & 0.1883 & 0.1201 & \underline{0.1923} & 0.1261 \\ \cmidrule(lr){2-14}
& SIM      & \textbf{0.1913} & \textbf{0.0237} & \underline{0.1917} & \textbf{0.0193} & \textbf{0.1916} & \textbf{0.0354} & \underline{0.1933} & \textbf{0.0598} & \underline{0.1889} & \textbf{0.0385} & \textbf{0.1937} & \textbf{0.0296} \\ \bottomrule

\end{tabular}}
\end{table*}

\subsection{Parameters Sensitivity Analysis}
\begin{figure}[h]
    \centering
    \begin{subfigure}[b]{0.95\textwidth}
        \centering
        \includegraphics[width=\textwidth]{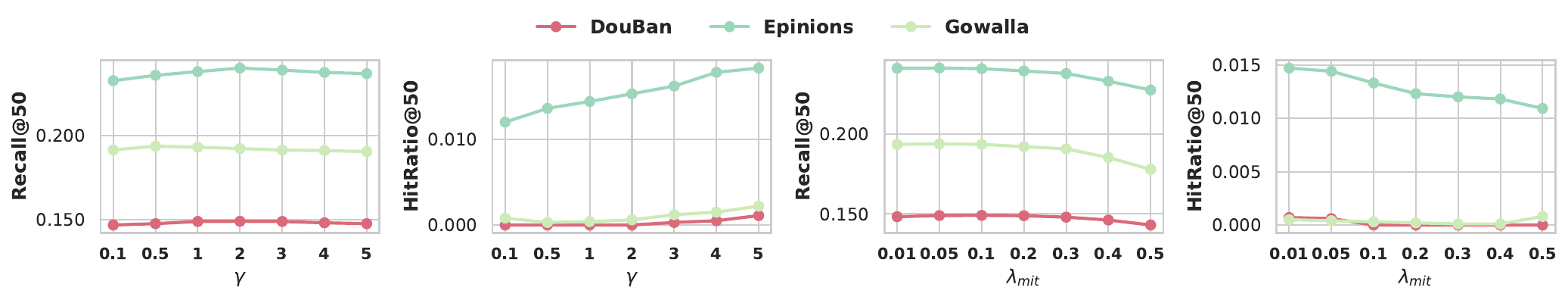}
        \caption{Parameter analysis of $\gamma$ and $\lambda_{mit}$ based on SSL4Rec.}
    \end{subfigure}    
    \hfill
    \begin{subfigure}[b]{0.95\textwidth}
        \centering
        \includegraphics[width=\textwidth]{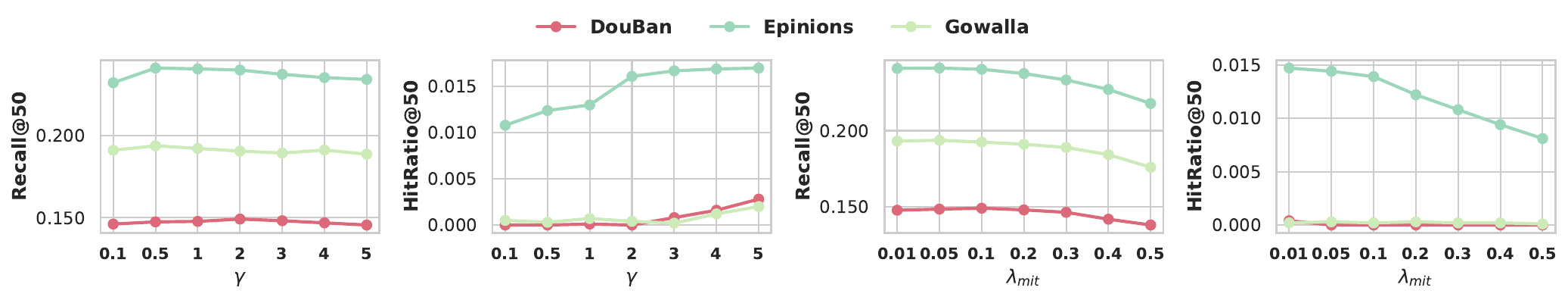}
        \caption{Parameter analysis of $\gamma$ and $\lambda_{mit}$ based on SGL.}
    \end{subfigure}
    \hfill
    \begin{subfigure}[b]{0.95\textwidth}
        \centering
        \includegraphics[width=\textwidth]{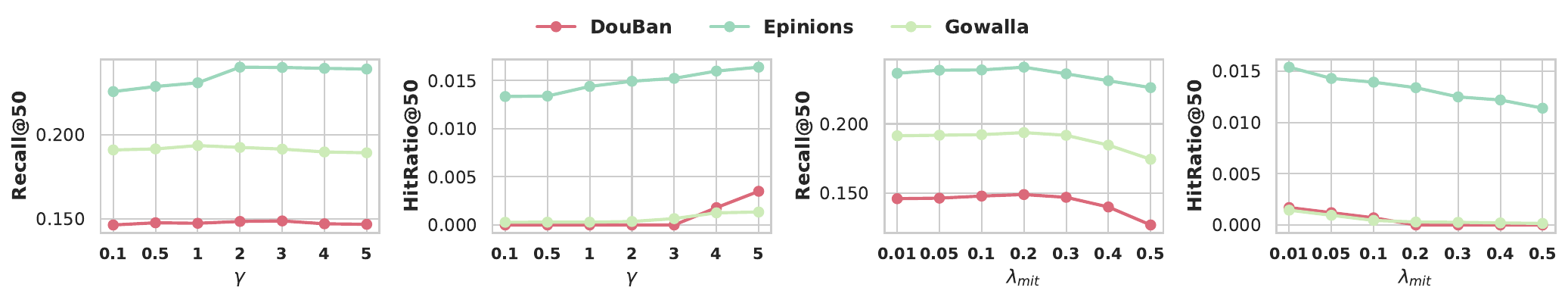}
        \caption{Parameter analysis of $\gamma$ and $\lambda_{mit}$ based on XSimGCL.}
    \end{subfigure}    \hfill
    \begin{subfigure}[b]{0.95\textwidth}
        \centering
        \includegraphics[width=\textwidth]{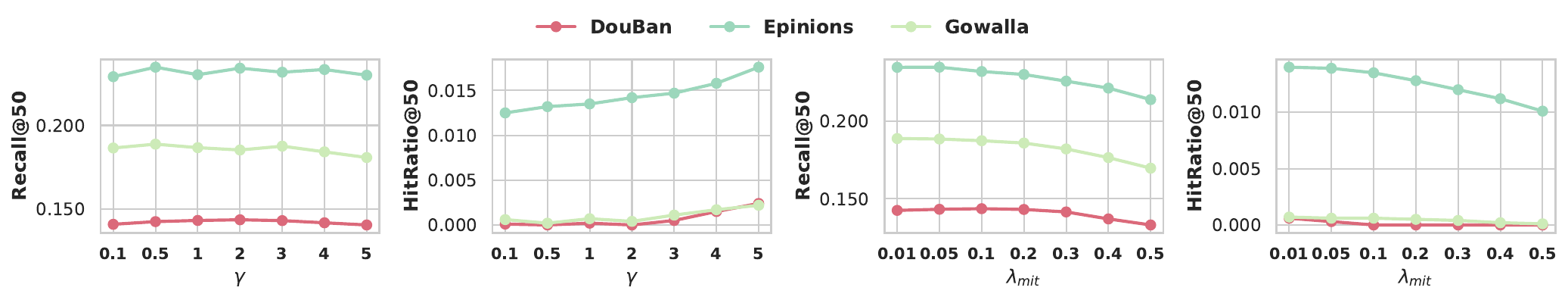}
        \caption{Parameter analysis of $\gamma$ and $\lambda_{mit}$ based on RecDCL.}
    \end{subfigure}
    \caption{Parameter analysis of $\gamma$ and $\lambda_{mit}$ based on different GCL-based recommendation models.}
    \label{parameter}
\end{figure}

We investigate the sensitivity of SIM to the hyperparameters $\gamma$ in Equation~(\ref{gamma}) and $\lambda_{mit}$ in Equation~(\ref{lambda}) on three datasets. The GCL-based recommendation model is set as XSimGCL, and the attack model is set as CLeaR. The experimental setup is as follows: we first fix $\lambda_{mit}$ at 0.1 and vary $\gamma$ within the set \{0.1, 0.5, 1, 2, 3, 4, 5\}; then, we fix $\gamma$ at 1 and vary $\lambda_{mit}$ within the set \{0.01, 0.05, 0.1, 0.2, 0.3, 0.4, 0.5\}. The results shown in Fig.~\ref{parameter} give us the following findings:
\begin{itemize}
    \item The experimental findings indicate that as $\gamma$ increases, the HitRatio also rises. We attribute this to the fact that a larger $\gamma$ reduces the number of candidate target items that can be identified, thereby decreasing the probability of successfully detecting true targets and ultimately weakening the defense effect. The inflection point for this trend occurs at $\gamma=3$ on the DouBan dataset and at $\gamma=1$ on the other two datasets.
    \item In terms of recommendation performance, it remains generally stable as $\gamma$ varies, suggesting that the effect of $L_{mit}$ is robust when $\lambda_{mit}$ is set appropriately. Notably, for the Epinions dataset, setting an extremely low value of $\gamma$ leads to a substantial decrease in recommendation accuracy. This suggests that Epinions is particularly sensitive to the precision of candidate target item identification; when too many incorrect items are flagged as candidates, the influence of $L_{mit}$ on recommendation performance becomes more pronounced. 
    \item  Regarding $\lambda_{mit}$, the results demonstrate that increasing its value enhances defense effectiveness, as evidenced by a lower HitRatio. However, an excessively large $\lambda_{mit}$ can cause a sharp decline in recommendation performance. When $\lambda_{mit}$ is relatively small (less than 0.2), recommendation quality remains stable across all three datasets.
\end{itemize}

\begin{summarybox}\textbf{Summary}:
A larger $\gamma$ weakens defense by missing more true targets; $\lambda_{mit}$ enhances defense but can harm recommendation quality if set too high. When $\lambda_{mit}<0.2$, SIM achieves stable accuracy and strong robustness across datasets.
\end{summarybox}

\subsection{Ablation Studies}
We also investigate the respective contributions of the two phases of SIM: anomaly detection and adversarial suppression. Since these two components are closely intertwined, we design two variants of SIM for ablation studies. The first variant, denoted as SIM w/o AS, retains the anomaly detection phase while discarding adversarial suppression. In this setting, we remove all candidate malicious items identified by anomaly detection, along with their associated interactions, and use the remaining data to retrain the model. The second variant, SIM w/o AD, removes anomaly detection and applies only adversarial suppression. Here, we randomly select 500 cold items as candidate target items and perform adversarial suppression accordingly. Under this experimental setup, we use XSimGCL as the GCL-based recommendation model and CLeaR as the attack method. The results across all three datasets are shown in Table \ref{ablation}, and we can get the following observations:
\begin{itemize}
    \item Notably, for SIM w/o AS, the HitRatio@50 (H@50) drops to zero, indicating that the candidate target items detected by Anomaly Detection completely cover the 10 actual target items. However, the brute-force removal of all candidate target items leads to a severe decline in recommendation performance. 
    \item In contrast, for SIM w/o AD, since it cannot directly locate the true target items, we observe that H@50 remains almost unchanged, yet the recommendation performance remains well. This suggests that separating cold items from their most similar user groups can unexpectedly enhance recommendation performance. 
\end{itemize}

\begin{table}[h]
\caption{Ablation study of SIM components.}
\label{ablation}
\centering
\resizebox{0.6\textwidth}{!}{
\begin{tabular}{@{}c|cc|cc|cc@{}}
\toprule
Dataset    & \multicolumn{2}{c|}{DouBan}        & \multicolumn{2}{c|}{Epinions}        & \multicolumn{2}{c}{Gowalla}          \\ \midrule
Metric     & \multicolumn{1}{c|}{R@50}   & H@50 & \multicolumn{1}{c|}{R@50}   & H@50   & \multicolumn{1}{c|}{R@50}   & H@50   \\ \midrule
SIM w/o AS & \multicolumn{1}{c|}{0.1289} & \textbf{0.0}  & \multicolumn{1}{c|}{0.2258} & \textbf{0.0}    & \multicolumn{1}{c|}{0.1746} & \textbf{0.0}    \\
SIM w/o AD & \multicolumn{1}{c|}{0.1483} & 0.3315 & \multicolumn{1}{c|}{0.2402} & 1.5294 & \multicolumn{1}{c|}{0.1924} & 0.1345 \\ 
\rowcolor{gray!15}
\textbf{SIM} & \textbf{0.1489} & \textbf{0.0} & \textbf{0.2410} & {1.4395} & \textbf{0.1937} & {0.0296} \\ \bottomrule
\end{tabular}}
\end{table}

\begin{summarybox}\textbf{Summary}:
Anomaly detection alone eliminates all targets but severely reduces recommendation quality. Adversarial suppression alone maintains accuracy but cannot effectively mitigate attacks. Both components are necessary for balanced defense and performance.
\end{summarybox}

\section{RELATED WORK}
\label{sec_related}
\subsection{GCL-Based Recommendation}
Graph Contrastive Learning (GCL)-based recommendation methods model recommender systems as graphs comprising user and item nodes interconnected through interactions. The core idea of GCL is to encourage similarity between representations of positive pairs while diminishing the similarity between negative pairs. This enables recommendation models to distill critical characters from complex user-item interactions.

Recent advancements have integrated various innovative augmentation strategies for GCL. For example, HHGR \cite{41zhang2021double} utilizes a dual-scale graph augmentation strategy tailored specifically for group-based recommendation scenarios. SSL4Rec \cite{55yao2021self} introduces random attribute and item masking techniques to generate distinct augmented views, focusing on maintaining consistency across diverse augmentations. Meanwhile, NCL \cite{43lin2022improving} develops a prototypical contrastive objective designed explicitly to capture contextual relationships between users/items and their corresponding neighborhood contexts. Additionally, semi-supervised paradigms have emerged with SEPT \cite{44yu2021socially} and COTREC \cite{xia2021self}, integrating social and session-based contexts to enhance recommendation outcomes.

Other notable approaches such as SGL \cite{08wu2021self} apply node and edge dropout techniques to create additional graph views explicitly for contrastive purposes. SimGCL \cite{09yu2022graph} directly generates augmented representations at the embedding level to simplify model training. Building upon SimGCL, XSimGCL \cite{25yu2022xsimgcl} further optimizes computational efficiency by streamlining both forward and backward propagation processes. RecDCL \cite{72zhang2024recdcl} extends the contrastive learning concept by incorporating feature-level considerations in addition to the traditional instance-level, enriching the representational capabilities of recommendation models.

\subsection{Targeted Promotion Attacks in Recommendation}
Targeted promotion attacks on recommender systems relied predominantly on heuristic manipulations of user behavior data. Early strategies, such as RandomAttack \cite{10lam2004shilling} and BandwagonAttack \cite{21li2016data} involved users assigning arbitrary ratings to various items or popular items. With the evolution of recommender systems becoming more robust, these simplistic methods have become less impactful, necessitating the development of advanced techniques employing sophisticated machine learning approaches \cite{62wang2021fast, 64yuan2023manipulating}.

As powerful generation mechanisms, generative adversarial networks (GAN) have stimulated various methods, including AUSH \cite{22lin2020attacking}, GOAT \cite{45wu2021ready}, and TrialAttack \cite{47wu2021triple}, which fabricate realistic fake user profiles derived from authentic user behavior patterns. These GAN-based attacks leverage discriminators trained to discern genuine from fabricated interactions, effectively guiding the adversarial generation process. Similarly, reinforcement learning strategies, exemplified by PoisonRec \cite{66song2020poisonrec}, LOKI \cite{46zhang2020practical}, and KGAttack \cite{48chen2022knowledge}, are designed to systematically construct deceptive interaction sequences, thus enhancing the targeted attack effectiveness.

Typically categorized as black-box attacks, these approaches operate under conditions where attackers lack explicit knowledge of the model's internal parameters. Nonetheless, certain advanced attacks assume partial or complete knowledge of the model. For instance, DLAttack \cite{59huang2021data} investigates scenarios where attackers possess detailed insights into the deep learning-based recommender model structures. AutoAttack \cite{73guo2023targeted} strategically crafts malicious profiles closely aligned with targeted user groups' preferences, ensuring precise influence while minimizing collateral damage. Similarly, GSPAttack \cite{67nguyen2023poisoning} explicitly targets graph-based recommender systems by manipulating limited malicious user interactions to increase the target item's prominence within recommendation lists. \textit{UBA}~\cite{79wang2024uplift} highlights that individual items have unique audience segments, posing varying levels of difficulty for targeted attacks. Consequently, employing a generalized strategy against all users indiscriminately can result in inefficient allocation of fake-user resources and reduced effectiveness of the attack.

\subsection{Defenses in Recommendation}
From the defender's standpoint, safeguarding recommender systems against promotion attacks typically involves two distinct but complementary strategies, each targeting a specific stage of the recommendation pipeline.

The initial line of defense emphasizes proactive detection mechanisms aimed at intercepting malicious inputs before they can impact the recommendation process. By identifying and isolating suspicious interactions early on, these mechanisms significantly mitigate potential threats, thus upholding the integrity of recommendation outcomes. For instance, \textit{GraphRfi}~\cite{65zhang2020gcn} integrates a fraud detection module directly into its recommendation architecture. This module evaluates user authenticity by dynamically modulating the influence of their ratings, effectively diminishing the contribution of potentially fraudulent users to the recommendation results. Similarly, \textit{NFGCN-TIA}~\cite{75wang2022detecting} constructs a specialized graph-based representation of user interactions, where nodes correspond to users and edges reflect their co-rating relationships. By selectively weighting and filtering out regular interactions, the resultant graph accentuates anomalous behaviors, enabling a three-layer GCN model to effectively identify malicious users. Extending this approach, \textit{DHAGCN}~\cite{78hao2023detection} further refines detection accuracy by extracting multiple user-centric features derived from item popularity trends and user rating patterns. These features inform a clustering strategy that groups users into distinct profiles, serving as high-quality inputs to enhance a GCN-driven detection model.

Acknowledging that no detection mechanism is foolproof, the second line of defense involves employing robust algorithms during the training phase of recommender models. These methodologies bolster system resilience by dynamically managing noisy inputs within the training data itself. For example, \textit{BOD}~\cite{71wang2023efficient} formulates recommendation denoising as a bi-level optimization task. Specifically, it utilizes gradient feedback from an inner denoising objective to strategically determine the weights for training samples in the outer optimization, thereby ensuring a cleaner learning environment. Furthermore, \textit{AutoDenoise}~\cite{77lin2023autodenoise} addresses fluctuating data distributions using a deep reinforcement learning approach. This method adaptively selects high-quality, noise-free data instances through an instance-denoising policy network, optimizing predictive performance dynamically. Moreover, \textit{DCF}~\cite{76he2024double} simultaneously tackles data sparsity and noise by implementing a dual-correction mechanism, effectively reassigning labels to both observed and unobserved interactions, thereby refining data quality and enhancing model robustness.

\section{Conclusion}
\label{sec_conclusion}
In this paper, we comprehensively analyzed and addressed a previously unrecognized vulnerability in GCL-based recommender systems: the increased susceptibility to targeted promotion attacks resulting from spectral smoothing induced by contrastive optimization. Through theoretical analysis and extensive empirical evaluations, we demonstrated that while GCL enhances recommendation robustness and alleviates popularity bias by promoting embedding dispersion, it also introduces an exploitable risk for targeted adversarial manipulation. Building on this insight, we developed CLeaR, a bi-level optimization framework that enables systematic investigation of this vulnerability and provides deeper understanding of the underlying mechanisms. To mitigate the risk without compromising the inherent strengths of GCL, we proposed SIM, a spectral irregularity mitigation framework capable of robustly detecting and suppressing anomalous embeddings. The experimental results validated the increased vulnerability of the GCL-based models to targeted promotion attacks and demonstrated the effectiveness of our proposed defense method in counteracting these risks.

Looking ahead, several promising directions remain to be explored. First, extending our spectral analysis framework to other recommendation paradigms, such as sequential or session-based recommenders, could reveal broader implications for contrastive learning. Besides, developing adaptive and computationally efficient spectral regularization techniques could further enhance the resilience of GCL-based models. 

\begin{acks}
We thank the anonymous reviewers for their insightful feedback and constructive comments. This work was supported by the National Natural Science Foundation of China (Grant No. 62176028) and the Australian Research Council under the streams of Discovery Early Career Research Award (Grant No. DE250100613), Future Fellowship (Grant No. FT210100624), Industrial Transformation Training Centre (Grant No. IC200100022), Discovery Project (Grant No. DP240101108), and Linkage Project (Grant No. LP230200892 and LP240200546). Co-author Ling Liu contributed to this project during unpaid weekends out of personal interests and
expresses gratitude to collaborators and family for their understanding. 
\end{acks}

\bibliographystyle{ACM-Reference-Format}
\bibliography{main}










\end{document}